\documentclass[aps,prd,notitlepage,superscriptaddress,preprintnumbers]{revtex4-1}

\usepackage{graphicx,xcolor}
\usepackage{physics}
\usepackage{amsmath,amssymb}
\usepackage{enumitem}   
\usepackage{dcolumn}
\usepackage[hidelinks]{hyperref}
\usepackage{cleveref}
\usepackage[normalem]{ulem}

\newcommand{\avg}[1]{\left\langle#1\right\rangle}
\newcommand{\ee}{e^+e^-}
\newcommand{\RP}{\{\mathrm{RP}\}}
\newcommand{\SP}[1]{\{\mathrm{SP}#1\}}
\newcommand{\bfq}{\mathbf{q}}
\newcommand{\GeV}{\mathrm{GeV}}
\newcommand{\fm}{\mathrm{fm}}

\begin{document}
\title{Distinguishing the sources of dielectron anisotropic flow at low beam energies}

\author{Renan Hirayama}
\affiliation{Helmholtz Research Academy Hesse (HFHF)\\GSI Helmholtz Center, Campus Frankfurt, Max-von-Laue-Str. 12, 60438 Frankfurt
am Main, Germany}
\affiliation{Institute for Theoretical Physics\\Goethe University, Max-von-Laue-Strasse 1,
60438 Frankfurt am Main, Germany}
\affiliation{Frankfurt Institute for Advanced Studies\\ Ruth-Moufang-Strasse 1, 60438 Frankfurt am Main, Germany}

\author{Hannah Elfner}
\affiliation{GSI Helmholtzzentrum für Schwerionenforschung\\Planckstraße 1, 64291 Darmstadt, Germany}
\affiliation{Helmholtz Research Academy Hesse (HFHF)\\GSI Helmholtz Center, Campus Frankfurt, Max-von-Laue-Str. 12, 60438 Frankfurt
am Main, Germany}
\affiliation{Institute for Theoretical Physics\\Goethe University, Max-von-Laue-Strasse 1,
60438 Frankfurt am Main, Germany}
\affiliation{Frankfurt Institute for Advanced Studies\\ Ruth-Moufang-Strasse 1, 60438 Frankfurt am Main, Germany}

\begin{abstract}
We present calculations of dielectron anisotropic flow in heavy-ion collisions at HADES beam energies from a hadronic transport approach. The ongoing experimental analysis employs the traditional reaction plane method to evaluate the flow coefficients $v_n$ and claims to see isotropic radiation from the thermal quark-gluon plasma. We show in this work, that in the region above the pion mass, the dilepton flow measurement might suffer from cancellation effects that mask the complicated underlying dynamics. Contributions from different baryonic and mesonic resonances show collective behaviour with different signs and lead to an overall vanishing elliptic flow. To differentiate the different contributions, we propose to employ the scalar product method, which exploits the previously measured hadronic flow to create different reference planes. As a proof of concept, we calculate the $v_2$ of dielectrons for Ag+Ag collisions at $\sqrt{s_{NN}}=2.55\ \GeV$ with both methods and investigate the contribution of each source, concluding that the scalar product method provides the proton and pion tagged flow coefficients as two distinct measurements, disentangling the various dilepton sources. \end{abstract}

\maketitle

\section{Introduction}\label{sec:intro}

Dileptons are known to be a good probe for the evolution of a heavy ion collision (HIC), because they do not couple strongly to the medium and leave it mostly unscathed, providing direct access to properties of the hot and dense matter from which they originate. With this power, however, comes the downside of accumulation: every stage of the collision produces electromagnetic radiation, so measurements are time-integrated. If an observable changes sign during the evolution, it will suffer from cancellation effects that make it harder to draw firm physical conclusions. 

One such observable that can take either positive or negative values is the flow coefficient $v_n$, which describes the transverse anisotropy of the system in momentum space. In particular, $n=2$ corresponds to the \emph{elliptic flow}, quantifying how much the system expands along one axis in comparison to its orthogonal direction in the transverse plane. A non zero hadronic $v_2$ has been used as a ``smoking gun'' for the presence of collectivity \cite{Ollitrault:1992bk,Luzum:2013yya}, and it imposes constraints on theoretical models. It is, however, a measurement on the final state hadrons, and therefore may not be sensitive to model details from before the freezeout \cite{Vujanovic:2019yih}. Since dileptons are produced throughout the entire evolution, they are more sensitive to the build-up of flow and can serve as a probe for model parameters, such as the relaxation time or a temperature-dependent shear viscosity at high beam energies \cite{Vujanovic:2016anq,Vujanovic:2017psb}; and may be useful to study the nuclear potentials and equation of state at low beam energies \cite{Reichert:2023eev}.

In the usual picture of elliptic flow, a positive (negative) $v_2$ corresponds to flow in (out of) the reaction plane, defined below. Hadrons in non-central HICs with a beam energy higher than $E_\mathrm{beam}\gtrsim5\ A\GeV$ generally exhibit positive elliptic flow, driven by the initial eccentricity of the resulting medium that creates a larger pressure gradient along the reaction plane. Below $E_\mathrm{beam}\lesssim0.1\ A\GeV$, the flow is also in-plane as the nuclei only bounce off each other, deflecting along the impact parameter direction. In the region between $0.15\sim3\ A\GeV$, though, it is out of plane due to the so called squeeze-out, in which the presence of spectators shadows the in-plane expansion \cite{Stock:2004iim}. This is the beam energy range where the HADES experiment at GSI operates \cite{HADES:2009aat}.

There are several methods to evaluate flow coefficients, the most basic being the \emph{Reaction Plane} (RP) method, in which the impact parameter $b$ serves as a reference axis from where the angles of final particles are defined. The plane spanned by the beam direction and the impact parameter is the reaction plane, tilted by an angle $\Psi_\mathrm{RP}$ with respect to the ``laboratory'' coordinate frame. In theory calculations $b$ is usually an input parameter, thus the RP method is easily applicable; in experimental analyses, however, $b$ is never directly accessible so the reaction plane must be estimated. Because of this difficulty, HIC experiments at high energies such as those performed at RHIC-BNL \cite{STAR:2022gki,sPHENIX:2024flow} and LHC \cite{CMS:2015xmx,ALICE:2022zks} have recently favored either the Scalar Product \cite{Voloshin:2008dg}, which measures the correlation between the flow of distinct sets of particles, or the multiparticle cumulant method \cite{Borghini:2001vi}, in which the flow signal is computed with azimuthal correlations between several particles. Equipped with the Forward Wall detector, HADES is able to detect the tranverse deflection of spectators, and thus construct a good estimate for the reaction plane \cite{Kardan:2018hna,HADES:2020lob}. 

While the RP method has proven very reliable for final state hadronic measurements, we argue in this paper that the same may not be true for dilepton observables. Specifically, the HADES collaboration has preliminary measurements of a $v_2\RP$ consistent with zero for dileptons of invariant mass $M_{ee}>0.12\ \GeV$, concluding that this proves dileptons can be used as penetrating probes \cite{Galatyuk:2020lvg, Schild:2024ywo,TalkWithGalatyukSchild}. This relies on a few assumptions, namely: dileptons leave the medium undisturbed; most of them are produced during the hot and dense stage; and there is little flow generated in this interval. Although the first two are correct, the last one is not. As we will show, the zero signal is due to cancellation effects between the in-plane flow of the dense stage with the out-of-plane flow of the late stage.

In HADES beam energies, the degrees of freedom are hadronic and the produced medium is dilute enough -- in comparison to higher collision energies -- to be described by kinetic transport. We employ the SMASH (Simulating Many Accelerated Strongly-interaction Hadrons) approach \cite{SMASH:2016zqf}, where the dilepton radiation is calculated directly from the hadrons in the simulation \cite{Staudenmaier:2017vtq}. This allows for the evaluation of event-by-event observables that naturally include final state effects, such as the aforementioned $v_2$ cancellations. In this work, we propose a complement to the ongoing analysis done by the HADES collaboration on the dielectron elliptic flow, by computing it with the Scalar Product method.

This paper is organized as follows: \cref{sec:SMASH} describes the hadronic transport approach used. Section \ref{sec:flow} shows the calculations with the reaction plane method for dielectron $v_2$, motivating the definition and usage of the scalar product method. The corresponding results are shown in \cref{sec:results} and discussed in \cref{sec:discussion}. Appendix \ref{sec:app:RP_predictions} contains predictions for the RP flow of dielectrons in Ag+Ag collisions at $\sqrt{s_{NN}}=2.55\ \GeV$ and \cref{sec:app:SP_pT} the transverse momentum dependent results from the scalar product method. 

\section{SMASH}\label{sec:SMASH}
We simulate HICs using the cascade mode of SMASH (\emph{Simulating Many Accelerated Strongly-interacting Hadrons}), a general purpose transport approach which evolves hadrons according to the relativistic Boltzmann equation \cite{SMASH:2016zqf}. The collision term consists of binary scatterings ($2\leftrightarrow2$) and resonance formation/decay ($2\leftrightarrow1$). To determine whether an interaction is possible, we employ a geometric collision criterion based on the total cross section. Hadronic decays have a rate equal to the inverse width in the hadron rest frame, given by a mass-dependent parametrization fixed to the vacuum value $\Gamma_0$ given by the Particle Data Group \cite{ParticleDataGroup:2018ovx} at the pole mass. The resulting channel is selected according to its branching ratio. Hadrons with $\Gamma_0<15$ keV (such as $\pi^0$ and $\eta$ mesons) are considered stable and do not decay mid-evolution, all other particles are treated as resonances with finite lifetimes.

The phase space distribution of hadrons in low beam energy collisions is well described by SMASH \cite{SMASH:2016zqf,Petersen:2018jag}, and the flow of protons and light nuclei was studied in \cite{Mohs:2020awg,Tarasovicova:2024isp} where mean-field nuclear potentials were used with moderate agreement to data. Due to the qualitative nature of this work, we do not include any potentials and leave an investigation on their effect to a future study.

Dilepton production in SMASH is limited to direct or Dalitz decays, with the specific channels for dilepton production described in detail in \cite{Staudenmaier:2017vtq}.
Since the electromagnetic branching ratio of resonances is very small, e.g. BR$_{\rho\to e^+e^-}\approx 10^{-5}$, an unfeasible number of events would be needed to accumulate statistics if we treated dilepton emission in the same way as the hadronic decays. Instead, we rely on a perturbative time-integration method, also known as \emph{shining} \cite{Heinz:1991fn}: at every time step with duration $\Delta t$ the resonance radiates its electromagnetic channels, but the emission carries the weight 
\begin{equation}\label{eq:shining_weight}
w_{R\to \text{e.m.}}(\Delta t)=\int\limits_0^{\Delta t}\frac{\dd{t}}{\gamma_R}\Gamma_{R\to \text{e.m.}}\left(M_{ll}\right),
\end{equation}
which is taken into account later when computing multiplicities. Here, $\gamma_R$ is the resonance's Lorentz factor, and $M_{ll}$ is the dilepton invariant mass. When the emission is direct ($R\to{l^+l^-}$), the latter is equal to the resonance mass, but in a Dalitz process ($R\to{Xl^+l^-}$) it is sampled uniformly from the available phase space. These perturbative decays are not \emph{actually} performed, in the sense of removing the decaying particle from the simulation, but the resonance propagates further until it is absorbed or decays hadronically. For hadrons considered stable, the dilepton decay is always performed, along with the proper phase space weight, at the end of the event. 

SMASH relies solely on vacuum properties, such that the resulting dilepton yield for $p+p$ collisions matches experimental data. However, the modifications caused by medium interactions in larger systems cannot be fully explained by the collisional broadening that arises in vacuum-based hadronic scatterings \cite{Hirayama:2022rur}, as is especially noticeable around the $\rho$ meson pole mass. If the shining method for vector mesons is replaced with their thermal rates \cite{Rapp:1999us,Rapp:2009yu}, the invariant mass spectrum becomes compatible to experimental data  \cite{Staudenmaier:2017vtq}. However, this requires coarse-graining the system evolution, where temperature and baryochemical potential profiles are extracted using an average over an ensemble of equivalent events, as has been done in \cite{Endres:2014zua,Endres:2015fna, Galatyuk:2015pkq, Seck:2020qbx}. This prevents the calculation of event-by-event observables, so we adhere to the shining method for vector mesons and make the caveat that the \emph{number} of dileptons in a given invariant mass is not perfectly compatible with HIC experiments. However, the qualitative behavior of the results in the following sections remains valid.

\section{Anisotropic flow}\label{sec:flow}
As described in the introduction, the azimuthal distribution of particles arriving in the detector after a collision experiment can be expanded in a Fourier series around the reaction plane angle \cite{Ollitrault:1992bk,Voloshin:1994mz,Luzum:2013yya}
\begin{equation}\label{eq:flow:fourier_expansion_azimuthal}
\dv{N}{\phi} = 1+2\sum_{n=1}^\infty v_n \cos [n(\phi -\Psi_\mathrm{RP})].
\end{equation}
Then, the flow coefficients $v_n$ can be computed as 
\begin{equation}\label{eq:flow:vnRP}
\begin{aligned}
v_n\RP&=\avg{\cos n (\phi-\Psi_\mathrm{RP})}\\&\equiv \dfrac{\displaystyle\int_{0}^{2\pi} \dd\phi\dfrac{\dd N}{\dd \mathbf{p}}\cos[n(\phi-\Psi_\mathrm{RP})]}{\displaystyle\int_{0}^{2\pi} \dd\phi\dfrac{\dd N}{\dd \mathbf{p}}}.
\end{aligned}
\end{equation}

In the HADES experiment, $\Psi_\mathrm{RP}$ is estimated from the deflection of spectators hitting the Forward Wall, as the impact parameter is not known. This is a well-established method for the flow of hadrons \cite{Kardan:2018hna,HADES:2020lob,Prozorov:2024wto}, and is also being used in the ongoing analysis of dielectron data \cite{Galatyuk:2020lvg,Schild:2024ywo}. By construction, the impact parameter in SMASH is aligned with the lab coordinate system, such that $\Psi_\mathrm{RP}=0$ and we compute the elliptic flow ($n=2$) as
\begin{equation}\label{eq:flow:elliptic}
v_2\RP=\avg{\frac{p_x^2-p_y^2}{p_x^2+p_y^2}}.
\end{equation}

\begin{figure}[ht]
\centering
\includegraphics[width=0.5\linewidth]{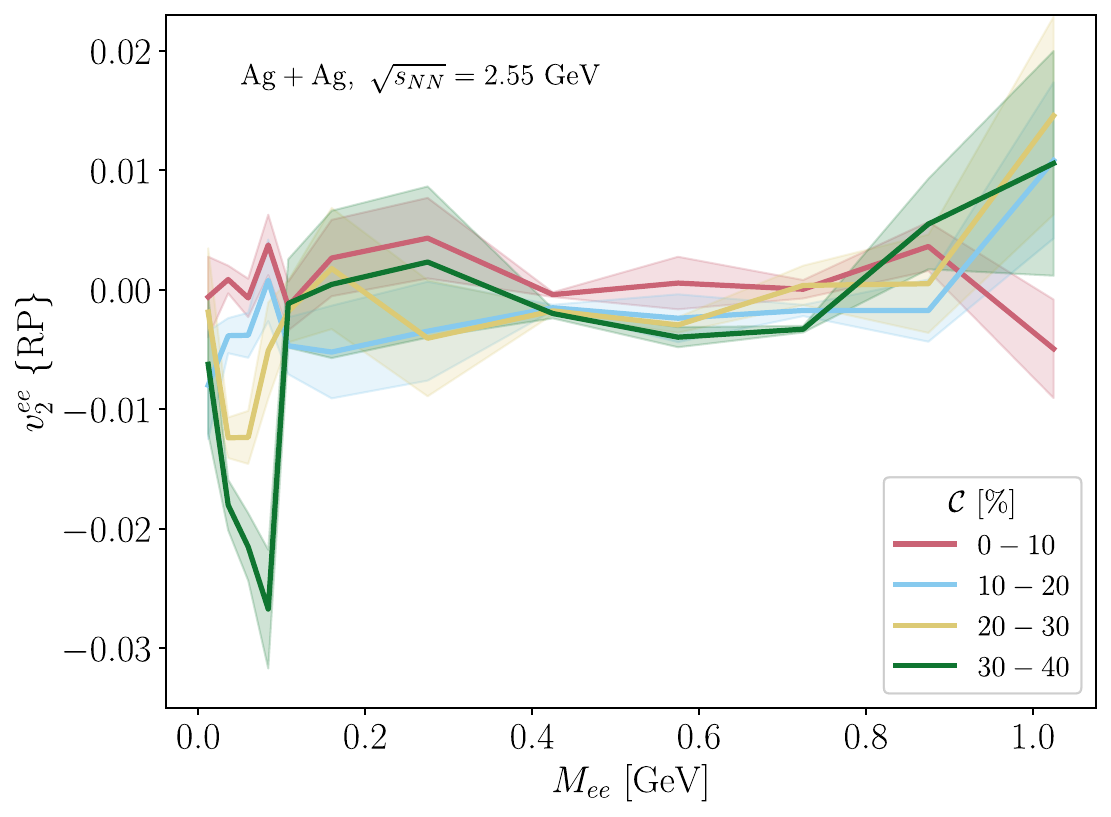}\caption{Elliptic flow of dielectrons, calculated with the reaction plane method \eqref{eq:flow:elliptic}, as a function of the invariant mass.}\label{flow:fig:RP_invmass}
\end{figure}

When applied for hadrons, the definition above can be evaluated either differentially in rapidity $y$ and transverse momentum $p_T$, or integrated over the phase space. Dileptons have an extra dimension in their phase space, namely the invariant mass $M_{ll}$ of the lepton pair. This degree of freedom separates low beam energy measurements of dielectrons in two regions: the \emph{pion region}, where $M_{ee}<m_\pi$ and the vast majority of dielectrons come from the Dalitz decay of $\pi^0\to\gamma e^+e^-$; above that is the \emph{resonance region}, where the largest contributions to the spectra are the Dalitz decays $\Delta\to N e^+e^-$ and $\eta\to\gamma e^+e^-$, as well as the direct decays of vector mesons $\rho,\omega,\phi\to e^+e^-$. In \cref{flow:fig:RP_invmass}, we show the dependence of the $v_2\RP$ on $M_{ee}$ for a Ag+Ag collision at HADES energies, calculated with the cascade mode of SMASH. 

The two invariant mass regions have distinct behaviors, with the low masses exhibiting out-of-plane flow, stronger for less central collisions, as expected for final state particles in this beam energy. Centrality selection has been performed by binning minimum bias events (0-40\% translated to impact parameter) by their final multiplicities in the bins corresponding to the experimental analysis. The resonance region has a faint negative flow between $0.4< M_{ee}<0.7 \ \GeV$, but the elliptic flow is mostly consistent with zero. These results are in qualitative agreement with preliminary results from the HADES collaboration \cite{Schild:2024ywo} but go against results in \cite{Reichert:2023eev} where a small but positive dilepton elliptic flow was calculated. We understand this tension as follows: in the latter work, radiation is computed by coarse graining, in which the events are ensemble-averaged to extract temperature and density fields used as input for thermal rates. Therefore the dilepton emission stops after $25\ \fm$, when the system is cold. However, dilepton-emitting resonances are still present in this dilute system, which now expands out-of-plane, and the corresponding late time emission may be enough to drop the accumulated elliptic flow to zero or even negative values.

\begin{figure}[ht]
\centering
\includegraphics[width=0.47\linewidth]{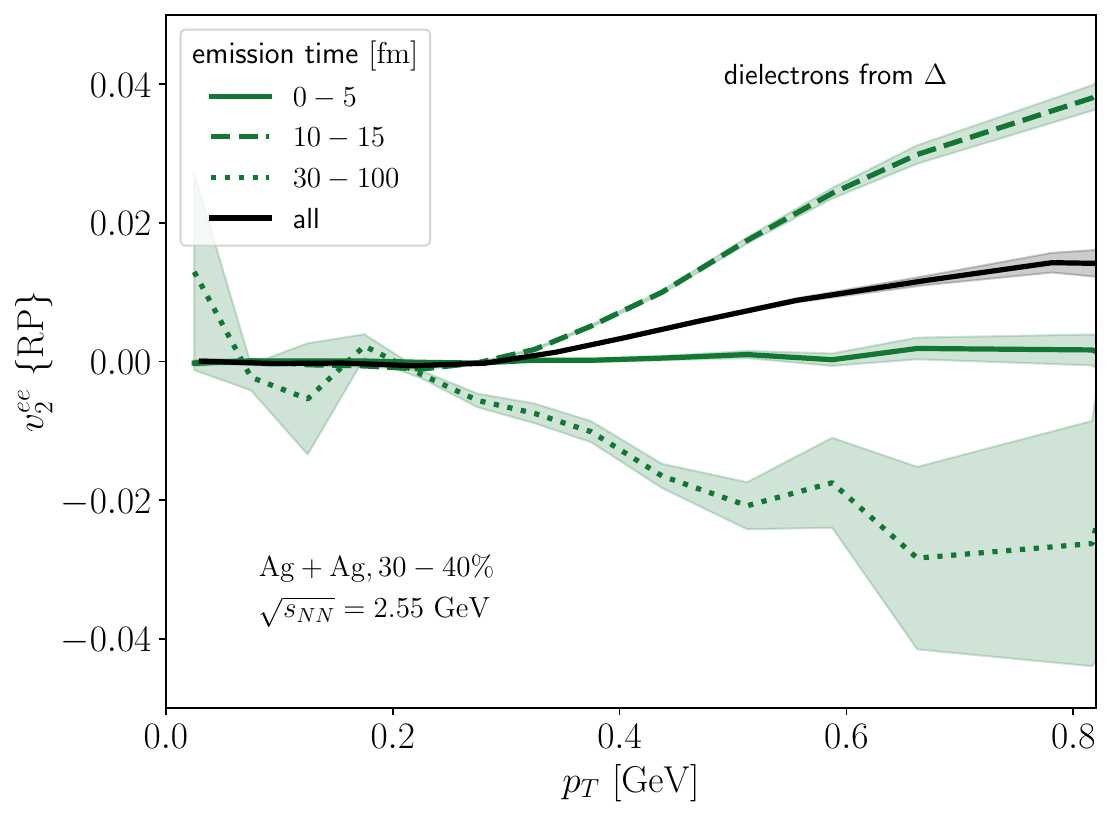}%
\hspace{0.02\linewidth}
\includegraphics[width=0.47\linewidth]{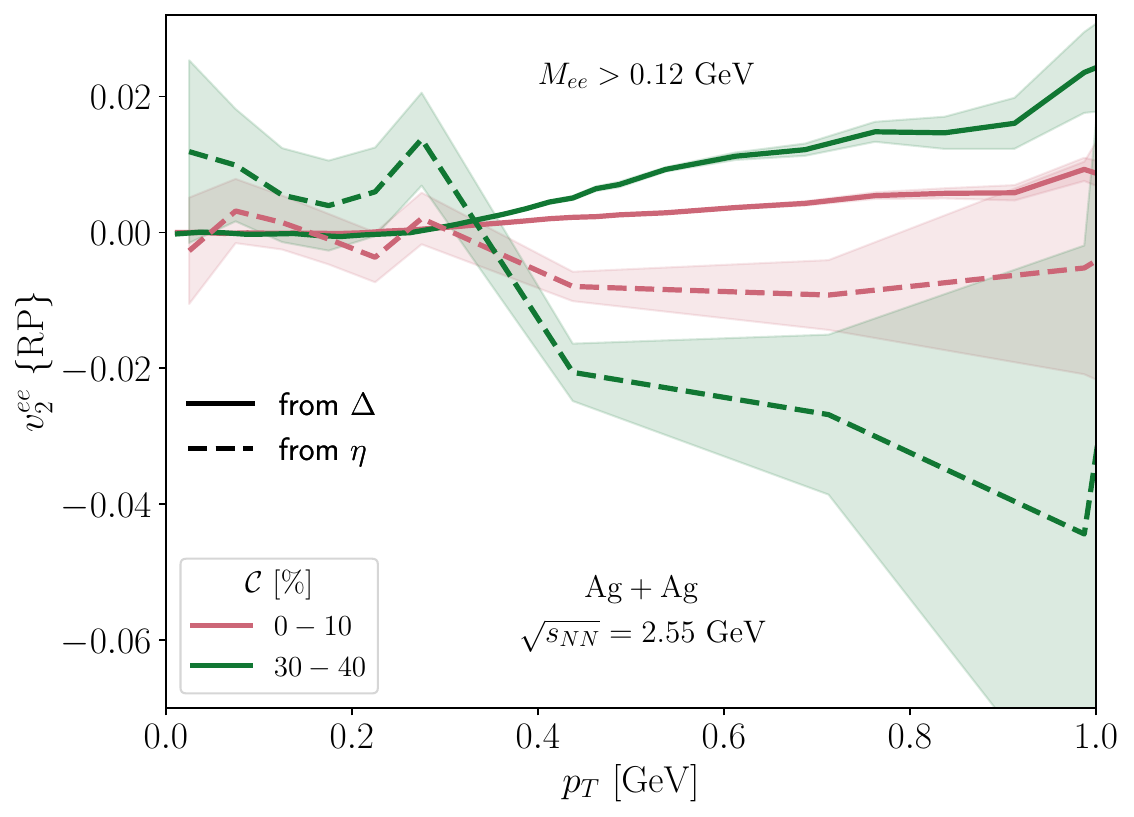}\caption{Competing contributions to dielectron $v_2\RP$. (Left) emissions from $\Delta$ baryons at different stages of the evolution and (right) time-integrated emissions from $\Delta$ and $\eta$.}\label{flow:fig:RP_contributions}
\end{figure}

These dynamics indicate that we must take care when analysing the phase space distribution of dileptons. As they probe the full evolution, not only the final state, the detected signal is a time integration over the flow generated during the whole evolution, meaning that opposing signals can cancel out. This is apparent in the left of \cref{flow:fig:RP_contributions}, where we show the flow \emph{only of dielectrons from $\Delta\to N\ee$}: in the interval $0-5\ \fm$, there has not been enough time to build up flow. This changes after $10\ \fm$, when the medium is hot and dense \cite{Hirayama:2022rur} and some in-plane expansion happens. As the system gets dilute and the spectators squeeze out the system, the late stage $\Delta$ acquires a negative $v_2$. The net result is a positive but small flow, less than half of that generated by the hot and dense phase. 

Another cancelation effect appears when considering the different particles that decay into dielectrons. In the right of \cref{flow:fig:RP_contributions}, we show two different and opposing processes that contribute to $v_2^{ee}$ in the resonance region: $\Delta\to N\ee$ as described above, is a resonance with overall positive flow. On the other hand, $\eta$ is considered stable enough to only decay into dielectrons at the late stage, when the system flows out-of-plane. Such mixing of sources with opposing flow patterns can explain the zero $v_2$ in the resonance region observed in HADES and in \cref{flow:fig:RP_invmass}, as the reaction plane method is unable to separate or weigh the contributions. This mixing also causes fluctuations in the measured signal, increasing the error bars in such a statistics demanding measurement. We provide further predictions for the differential $v_2\RP$ in \cref{sec:app:RP_predictions} that might be useful to compare with future measurements of the HADES collaboration.

In order to disentangle the different dilepton sources, we propose studying the anisotropy of dileptons based on the existing knowledge of the hadronic flow. To do so, we first define the event flow vector of species $X$
\begin{equation}\label{eq:flow:qn}
\bfq^X_n(\Omega) = \dfrac{\displaystyle\int_\Omega\dd{\Omega}\int_0^{2\pi} \dd\phi\dfrac{\ \dd N^X}{\dd\Omega\dd\phi} e^{in\phi}}{\displaystyle\int_\Omega\dd{\Omega}\int_0^{2\pi} \dd\phi\dfrac{\ \dd N^X}{\dd\Omega\dd\phi}},\end{equation}
for a region $\Omega$ of the phase space. In a given event, this vector represents the largest $n$-order momentum anisotropy of $X$, so the real part of its event-average is equivalent to \eqref{eq:flow:vnRP} if there are no multiplicity fluctuations. We then construct $v_2$ as the covariance between the event flow vector of dileptons and the direction of the hadron flow:
\begin{equation}\label{eq:flow:v2EP}
\begin{aligned}
v_n^{ll}\{\mathrm{EP}|h\}(\Omega)&\equiv\mathrm{Cov}\left[\bfq_n^{ll}(\Omega),\frac{\bfq_n^h}{|\bfq_n^h|}\right] \\&= \avg{\bfq_n^{ll}(\Omega)\cdot \frac{\bfq_n^{h\dagger}}{|\bfq_n^h|}} - \Big\langle\bfq^{ll}_n(\Omega)\Big\rangle\cdot\avg{\frac{\bfq^{h\dagger}_n}{|\bfq^h_n|}},
\end{aligned}
\end{equation}
where the averages over individual flow vectors vanish by symmetry. This is the \emph{Event Plane} method, in which the hadronic event flow $\bfq^{h}$ is taken over the full phase space, while dileptons are taken differentially in $\Omega$. If the strength of the hadronic reference flow fluctuates too much, an event with a small hadronic flow would contribute as much as a very anisotropic one. To prevent this, its normalization should be replaced by the root mean square, defining
\begin{equation}\label{eq:flow:v2SP}
v_n^{ll}\{\mathrm{SP}|h\}(\Omega)= \frac{\avg{\bfq_n^{ll}(\Omega)\cdot\bfq_n^{h\dagger}}}{\sqrt{\avg{|\bfq_n^h|^2}}},
\end{equation}	
which will favor the events with a larger reference anisotropy. This is the \emph{Scalar Product} method. We checked that, in all of our results, the definitions \eqref{eq:flow:v2EP} and \eqref{eq:flow:v2SP} only differ by a small margin, which may not be the case in experiment due to the limited acceptance and systematic uncertainties. Therefore we adhere to the low-resolution scalar product method in what follows. The definitions above were suggested in \cite{Paquet:2015lta,Vujanovic:2016anq} for RHIC and LHC energies, but we here we propose it for HADES energies, as it leads to fewer negative contributions, and hence carries less cancellations.
	
One further information accessible by the HADES detector is the particle identification, which we can exploit to construct different reference flow vectors. We nicknamed this choice as \emph{tagging}. Depending on the species $h$ chosen, the scalar product in the numerator of \eqref{eq:flow:v2SP} will highlight the channels that correlate strongly to $h$, and thus disentangle the different sources. For example, the $p$-tagged flow $v_n^{ee}\SP{|p}$ will have a large contribution from $\Delta$ baryons, because of the Dalitz decay $\Delta\to N\ee$. On the other hand, there is no direct link between $\eta$ mesons and protons, so their contribution to this quantity will be small. 

\section{Results with the Scalar Product method}\label{sec:results}

We simulate Ag+Ag collisions at $\sqrt{s_{NN}}=2.55\ \GeV$ ($E_\mathrm{kin}=1.58\ A\GeV$) with impact parameters ranging between $0$ and $7.7\ \mathrm{fm}$, corresponding to $0-40\%$ in Glauber model calculations \cite{nuclearoverlap}. These are then further divided in subclasses of 10\% width by multiplicity. We choose the reference at midrapidity, $|y-y_\mathrm{CM}|<0.25$, selecting protons with $p_T>0.3\ \GeV$ and pions with $p_T>0.1\ \GeV$, as this region contains enough particles and the hadronic flow is non-zero. Following the ongoing HADES analysis, we apply a cut in dilepton opening angle $\alpha_{ee}>0.9^\circ$ and in the single lepton momentum $|p_e|>0.1\ \GeV$. The results shown in this section take dielectrons at midrapidity, but are in principle applicable to the forward region as well.

\subsection{Time evolution}\label{sec:results:time_ev}

In order to make sense of definition \eqref{eq:flow:v2SP}, we first investigate how the contribution of separate sources changes over time, compared to the final hadron flow. Naturally, this can only be done in a theory calculation, but demonstrates the effect of choosing different tags. Because  we observed the same behavior in both pion and resonance mass regions, there is no cut in $M_{ee}$ in this subsection.

\begin{figure}[ht]
\centering
\includegraphics[width=0.47\linewidth]{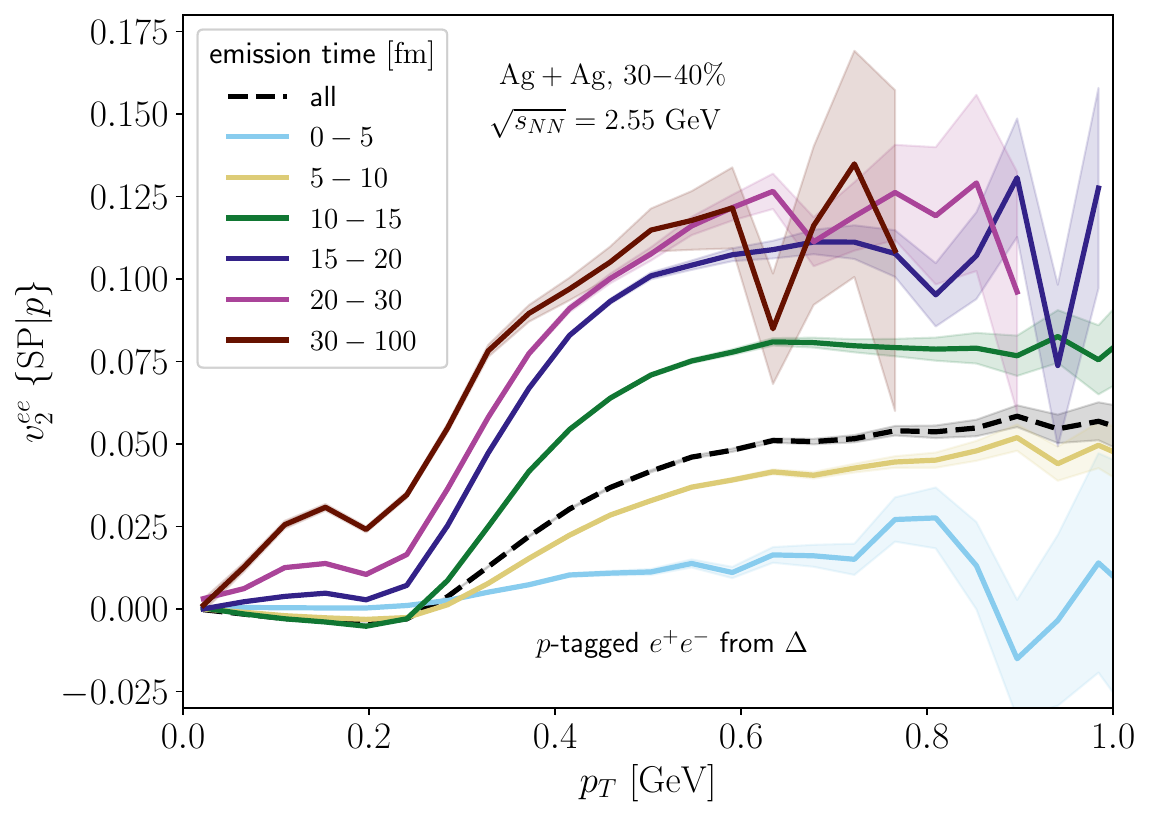}%
\includegraphics[width=0.47\linewidth]{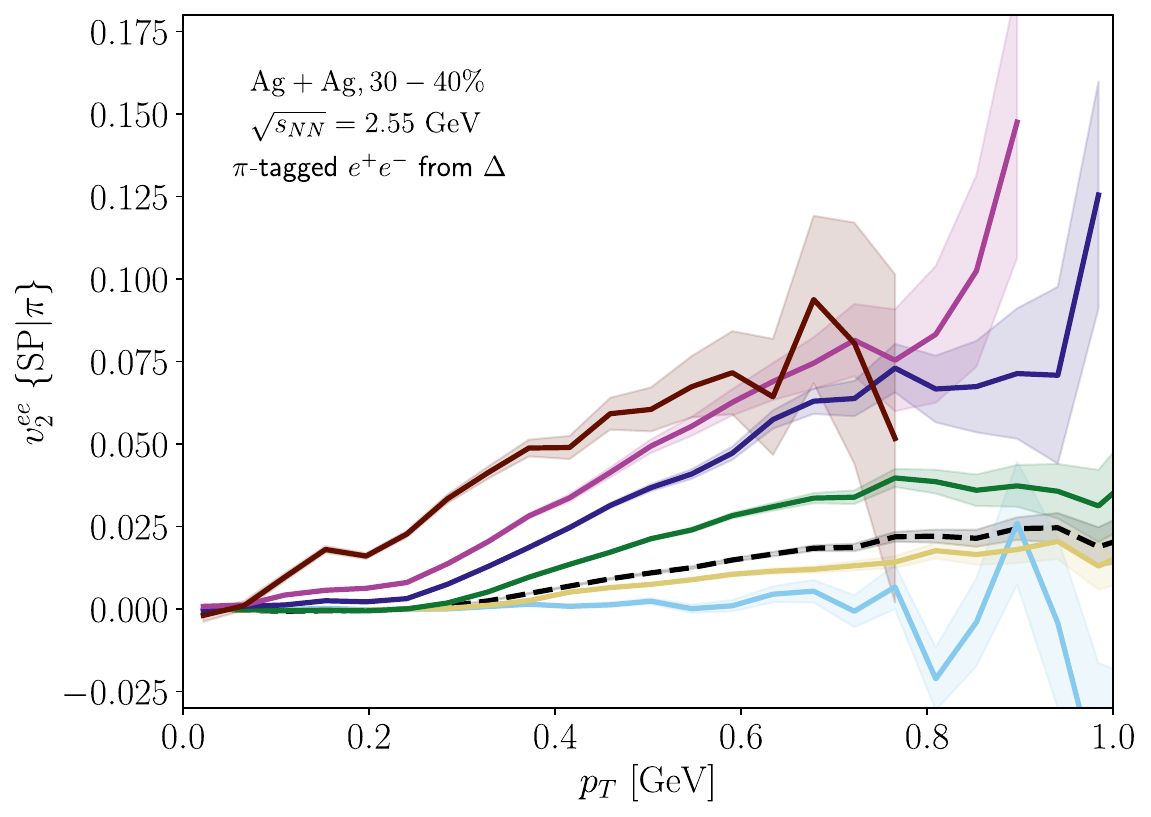}

\includegraphics[width=0.47\linewidth]{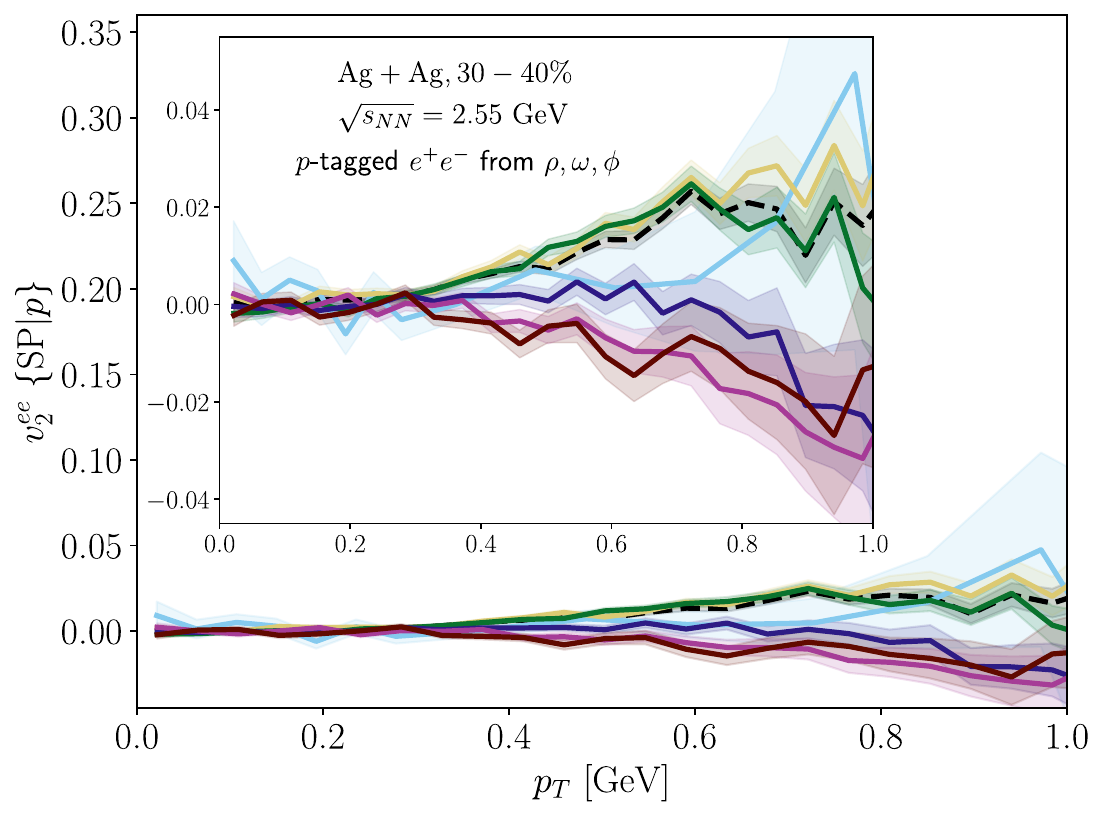}%
\includegraphics[width=0.47\linewidth]{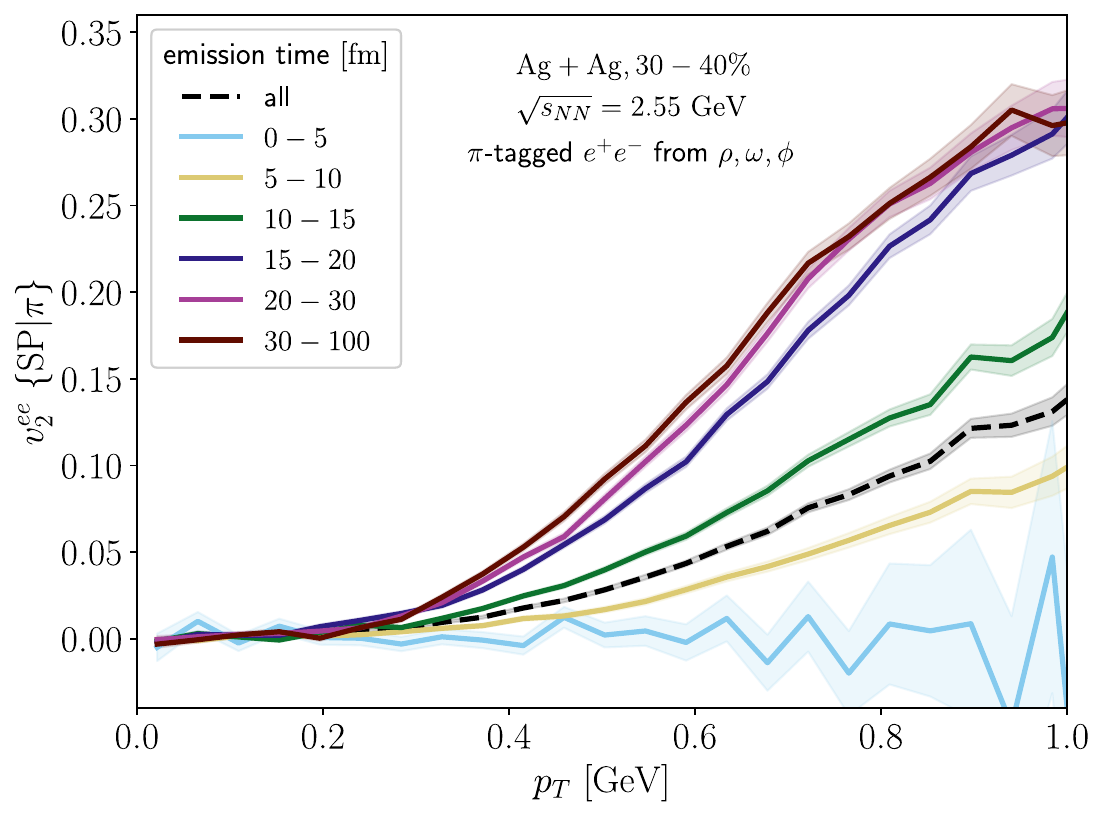}
\caption{Scalar product flow of dielectrons radiated by (upper) $\Delta$ baryons and (lower) vector mesons. Solid lines show the flow signal at different time intervals, and the dashed black lines show the full time integrated contribution of the source. The reference plane is constructed with (left) protons or (right) pions.}\label{flow:fig:SP_pT_time} 
\end{figure}

The upper plots of \cref{flow:fig:SP_pT_time} show the evolution of the $v_2^{ee}$ from the decay $\Delta\to N\ee$, tagged with protons on the left plot and pions on the right. In both tags, we see that the initial $\Delta$ contribution is almost zero, but increases with the expansion of the system until around $15-20\ \fm$. After that, the curves shift to the left as the transverse expansion slows down; at low $p_T$ the late stage signal has a bump. This happens because the $\Delta$ has a large elastic cross section with both baryons and mesons at low interaction energies, so they lose momentum while maintaining the average direction of motion -- and consequently, the anisotropy. The integrated signal follows the emssion of some time between $5-15$ fm, consistent with the coarse-grained result in \cite{Reichert:2023eev}. Unlike in the reaction plane method shown on the left of \cref{flow:fig:RP_contributions}, the scalar product signal does not change sign mid-evolution, and the emission is always aligned with the final state. At all times, the $p$-tagged signal is stronger than the $\pi$-tagged. This happens because the $\Delta$ is a more important source of protons than of pions, as the latter can be produced by many resonances and is more easily absorbed. Accordingly, their dilepton emission is more correlated to the former. Finally, we note that the $0-5$ fm contribution of the $p$-tag grows marginally as a function of $p_T$, which we believe is due to some decayed protons of these initial $\Delta$ surviving the evolution. This would not be the case for the $\pi$-tag because the corresponding pions would -- as before -- be more easily absorbed.

The lower plots of \cref{flow:fig:SP_pT_time} show the evolution of $v_2^{ee}$ from vector mesons ($\rho$, $\omega$, and $\phi$) calculated with the scalar product method \eqref{eq:flow:v2SP}. Unlike the $\Delta$ baryons, there is little connection between them and protons, so their contribution to $v_2^{ee}\SP{|p}$, shown in the left plot, is close to zero. On the other hand, they decay directly into pions, so they contribute much more to $v_2^{ee}\SP{|\pi}$, as we show in the right plot. Amidst the dense phase, both tags see the radiation emitted in alignment with the final state, with a positive signal, which suggests that hadrons are moving collectively. Towards the dilute stage, however, the $p$-tagged $v_2^{ee}$ changes sign -- remaining small -- while the $\pi$-tagged keeps increasing. We believe that this is due to the shadowing caused by protons on the vector mesons, favoring the production of these resonances in the orthogonal direction. Since the elastic cross section between these mesons and the surrounding cold medium is small, we only see a minor shift to lower momentum in the final stages. Similar to the $\Delta$ baryons, the time-integrated emission in both tags is also representative of the contributions between $5-15$ fm.

We present the overall $p_T$ dependence of the elliptic flow analysed with the hadron tagged scalar product method, including all sources, in \cref{sec:app:SP_pT}.

\subsection{Invariant mass}\label{sec:results:inv_mass}

\begin{figure}[ht]
\centering
\includegraphics[width=0.47\linewidth]{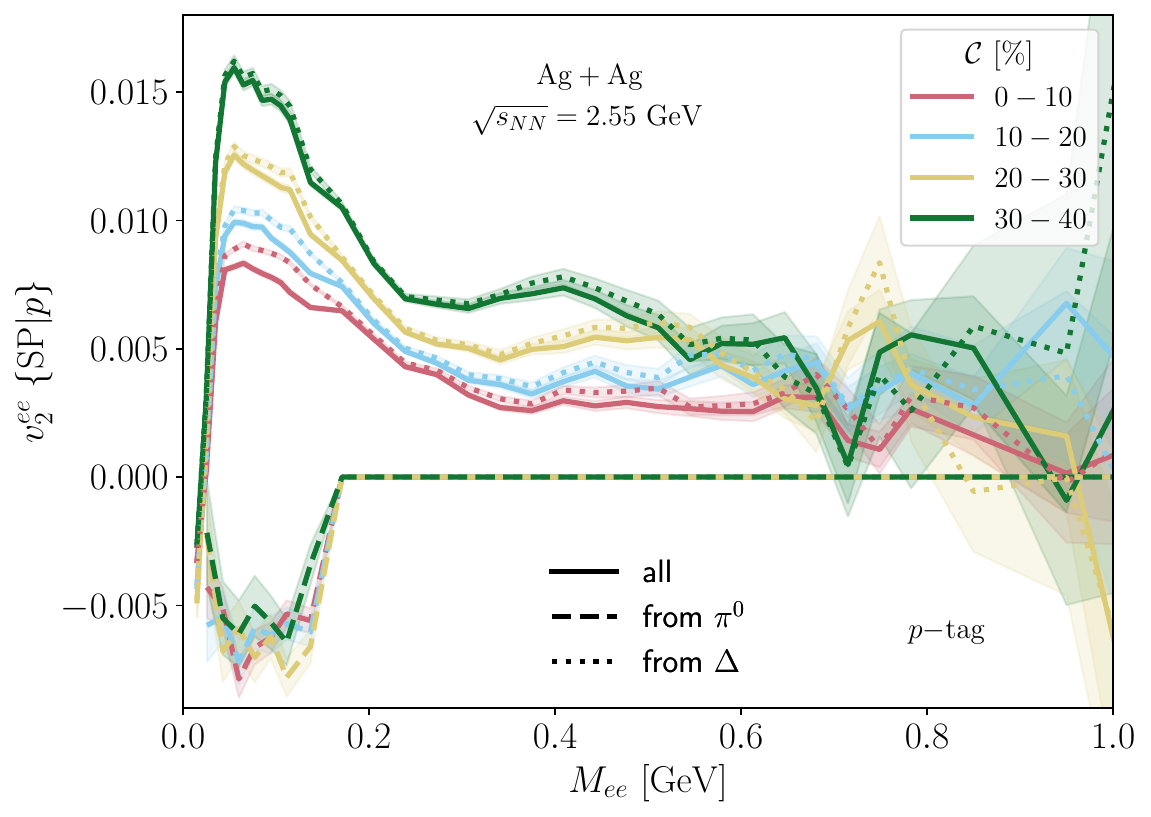}%
\hspace{0.02\linewidth}
\includegraphics[width=0.47\linewidth]{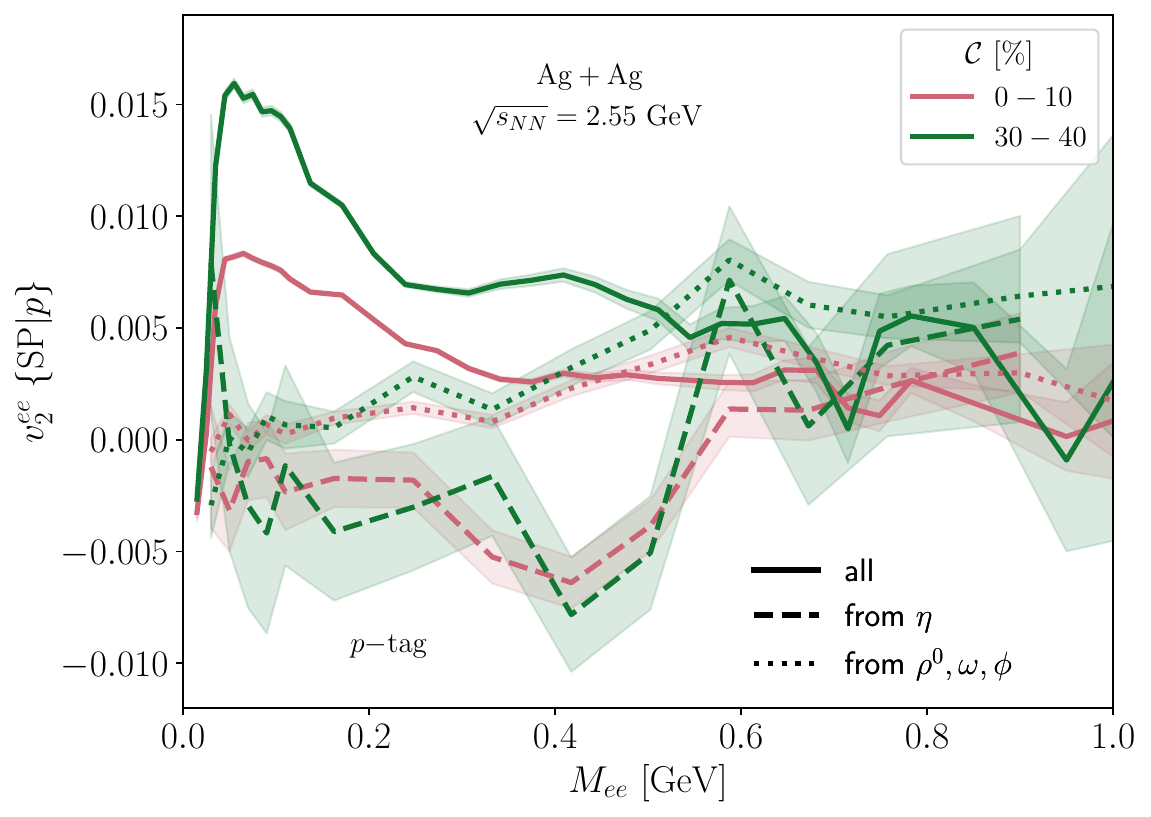}
\caption{Sources of proton-tagged dielectron $v_2$ with the scalar product method as a function of invariant mass, from decays of (left) $\pi^0$ and $\Delta$; (right) $\eta$ and vector mesons.}\label{flow:fig:SP_invmass_tag-proton}
\end{figure}

\begin{figure}[ht]
\centering
\includegraphics[width=0.47\linewidth]{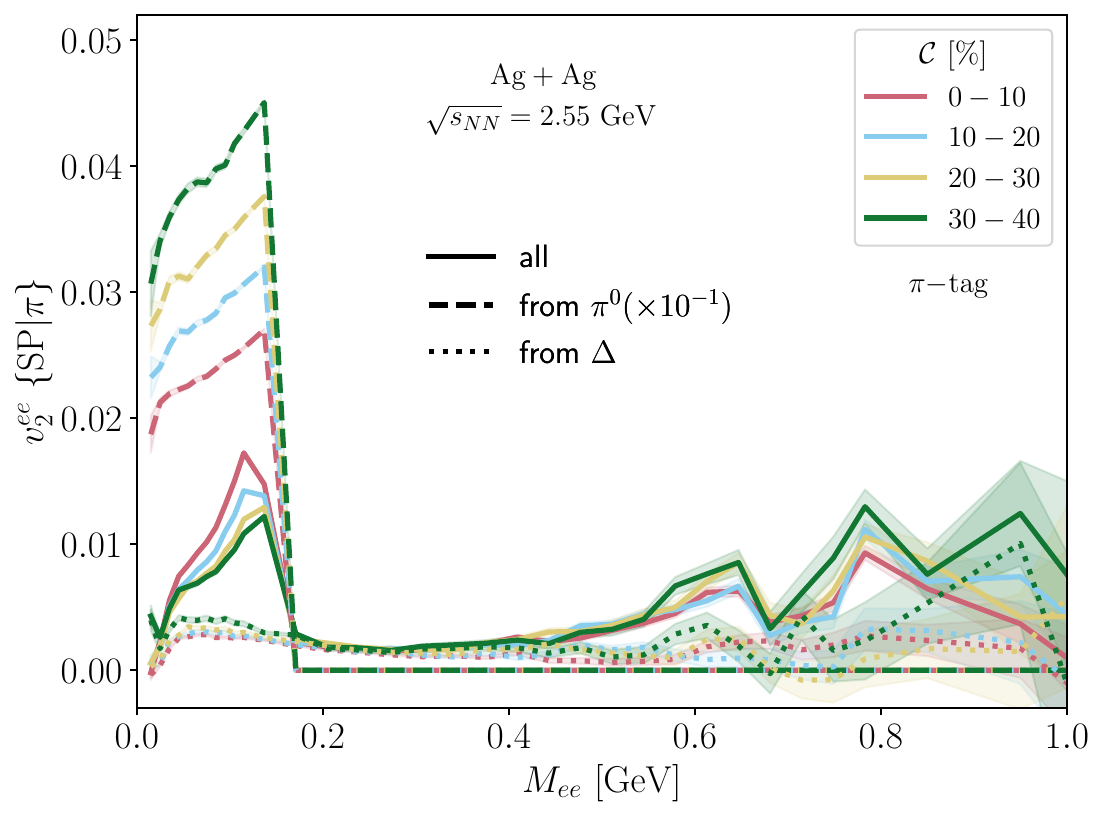}%
\hspace{0.02\linewidth}
\includegraphics[width=0.47\linewidth]{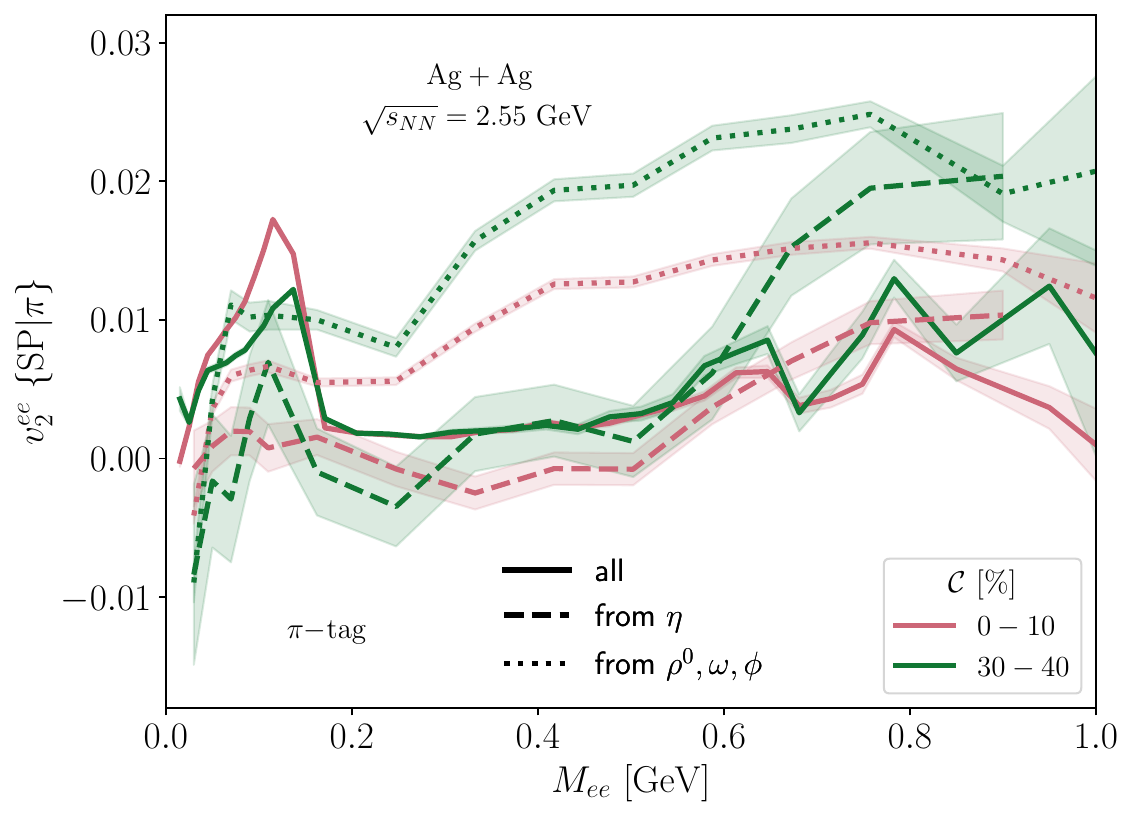}
\caption{Same as \ref{flow:fig:SP_invmass_tag-proton}, but using the charged pions as reference flow. The $\pi^0$ contribution is scaled by $0.1$ for visualization.}\label{flow:fig:SP_invmass_tag-pion}
\end{figure}

Now we study how the different sources contribute to a combined $v_2^{ee}\SP{}$, including the final-state decays $\pi^0\to\gamma\ee$ and $\eta\to\gamma\ee$. Figs. \ref{flow:fig:SP_invmass_tag-proton} and \ref{flow:fig:SP_invmass_tag-pion} show the $p$- and $\pi$-tagged signals respectively as a function of the dielectron invariant mass for different centrality classes, with the phase space cuts mentioned in the beginning of \cref{sec:results}. In some plots, we omit certain centrality classes for a better visualization, but they obey the general trend described in the text. In what follows, colors denote centrality classes and the line styles mark the different sources, with the solid line showing the total and measurable signal.

We immediately see in \cref{flow:fig:SP_invmass_tag-proton} that $v_2^{ee}\SP{|p}$ is a good predictor of the single contribution from $\Delta\to N\ee$, while other sources do not have a strong influence: it is peaked at low $M_{ee}$ and larger for less central collisions up to $M_{ee}\sim0.6\ \GeV$; consequently the same is true for the full signal. The $\pi^0$ contribution is negative but without a significant centrality dependence. Naturally, it is only present below the pion mass, reducing the total signal by a small amount from the $\Delta$ contribution in this region. The $\eta$ curves are consistent with 0 for low masses, slightly negative between $0.3<M_{ee}<0.5\ \GeV$ and slightly positive above that, but without a monotonic dependence on centrality. The average vector meson radiation is aligned with the final state protons, as we showed in \cref{flow:fig:SP_pT_time}, and roughly increases with dielectron mass.

Tagging the dielectron flow with pions results in a very different signal, as shown in Fig \ref{flow:fig:SP_invmass_tag-pion}. The radiation from $\pi^0$ predictably has the largest flow for $M_{ee}<m_\pi$, between $0.2$ to $0.45$, while the $\Delta$ reaches at most $0.005$ in the same region. Even though both behave intuitively with centrality -- with stronger flow in more peripheral collisions --, the interplay of weights in different classes results in a larger overall signal for central events, where pion production is augmented. In the resonance region, the $\Delta$ only coincides with the full emission between $m_\pi<M_{ee}\lesssim0.4\ \GeV$, where it is very small and exhibits no centrality dependence. The vector meson contribution is always positive with the expected centrality dependence; the same is true for the $\eta$ curve above $M_{ee}\gtrsim0.5\ \GeV$. Because of them, the average signal is above the $\Delta$ contribution, with a structure of peaks around $M_{ee}\sim0.6$ and $0.8\ \GeV$ common to all centralities.

\section{Discussion and outlook}\label{sec:discussion}

While the reaction plane method is well established for the measurement of flow anisotropy of hadrons at HADES, the same is not yet true when it comes to dielectrons. This is because the electromagnetic radiation comes from all stages of the fireball expansion and from various hadronic sources, so its phase space distribution does not necessarily correlate well with the reaction plane (or spectator deflection). We have shown that cancellations take place both due to the time-integrated nature of this observable, as well as the uncorrelated contributions from various dilepton sources. This means that the zero flow preliminarily observed by the HADES experiment might not result from an isotropic evolution of the hot and dense medium, but rather from non-trivial late stage cancellations.

In this manuscript, we proposed using the scalar product method \eqref{eq:flow:qn},\eqref{eq:flow:v2SP} to complement the ongoing analysis, exploiting the readily available measurements of hadronic flow to construct different reference planes, which highlights specific contributions. By \emph{tagging} the dileptons with protons or pions, two distinct flow measurements become available ``for free'', without the need for further data taking. This method carries smaller cancellation effects, since sources without a direct link to the reference hadron have a contribution close to zero to the total signal, instead of a negative one. In most regions of the phase space, this analysis leads to the expected centrality dependence, with a larger flow for more peripheral collisions.

Generally, the proton-tagged quantity $v_2^{ee}\SP{|p}$ mimics the flow of the radiation from $\Delta$ baryons, suggesting that it can be a good probe for characterizing the nuclear potentials which affect production and dynamics of this source. Because all sources of dileptons are also sources of pions (plus the pion itself), $v_2^{ee}\SP{|\pi}$ is a mix of mostly positive contributions, leading to a non trivial centrality dependence at low invariant mass. We show in \cref{sec:app:SP_pT} that this arises in the low $p_T$ region, suggesting that it is a non-flow effect caused by slow pions created by the decay of resonances in the dilute stage. We also see a non-zero signal for $M_{ee}>0.4\ \GeV$ caused by $\eta$ and vector mesons. In both tags, the time-integrated radiation of short-lived resonances is representative of the radiation from the hot and dense evolution between $5-15\ \fm$. The above results show that dileptons can indeed serve as penetrating probes, provided the construction of an observable that suppresses the other contributions. 

We only applied to this analysis the cuts described in the beginning of sec. \ref{sec:results}, but given that the flow signal increases with $p_T$ (see the appendices), using weights for the event flow vector \eqref{eq:flow:qn} or applying a further cut of a minimum $p_T$ would enhance it, and might be useful to remove some non-flow effects. Finally, we note that the calculations in this work serve as a proof of concept and are not quantitative, as only the cascade mode of SMASH is used. A study on the effect of mean-field potentials is left for the future, since this imposes a larger computational cost and therefore is not easily achievable in event-by-event calculations.

\section{Acknowledgements}

We thank Tetyana Galatyuk and Niklas Schild for providing insights on the status and plans of HADES. This work was supported by the Helmholtz Forschungsakademie Hessen für FAIR (HFHF). We also acknowledge the support by the State of Hesse within the Research Cluster ELEMENTS (Project ID 500/10.006), and by the Deutsche Forschungsgemeinschaft (DFG, German Research Foundation) – Project number 315477589 – TRR 211. Computational resources have been provided by the GreenCube at GSI. 

\appendix
\section{Reaction Plane predictions}\label{sec:app:RP_predictions}

\begin{figure}[h!]
\centering
\includegraphics[width=0.333\linewidth]{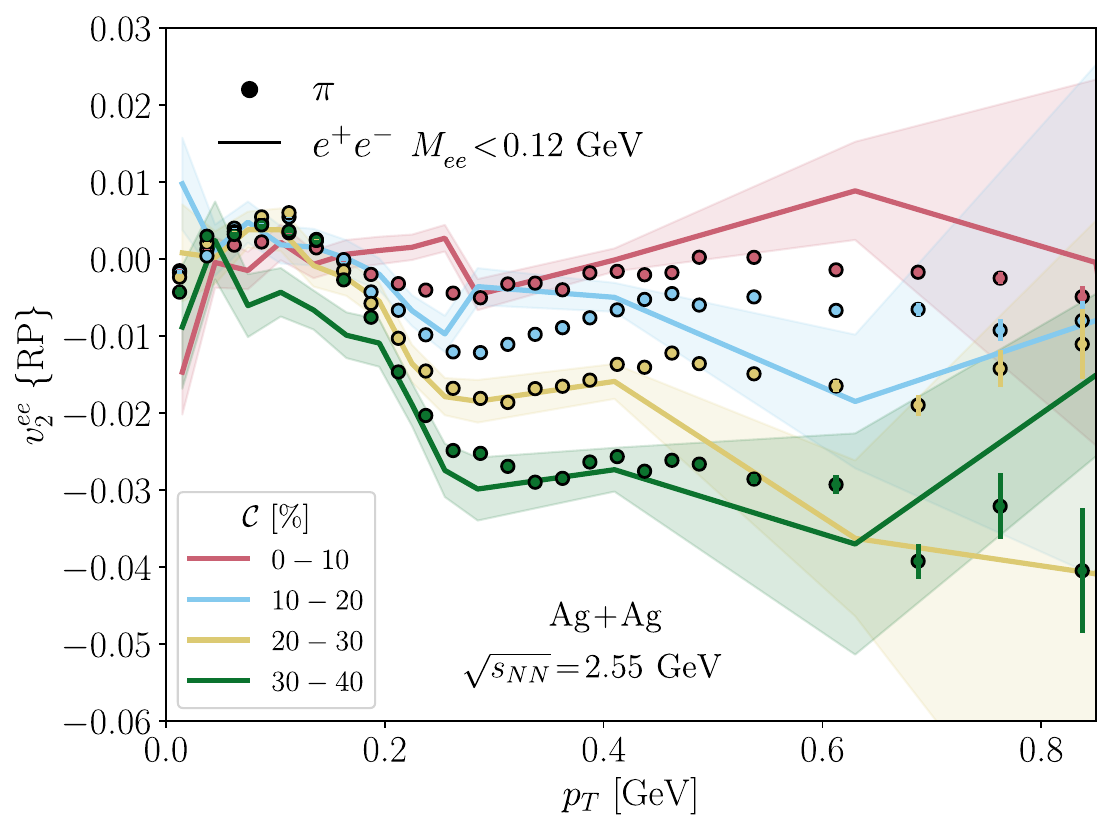}%
\includegraphics[width=0.333\linewidth]{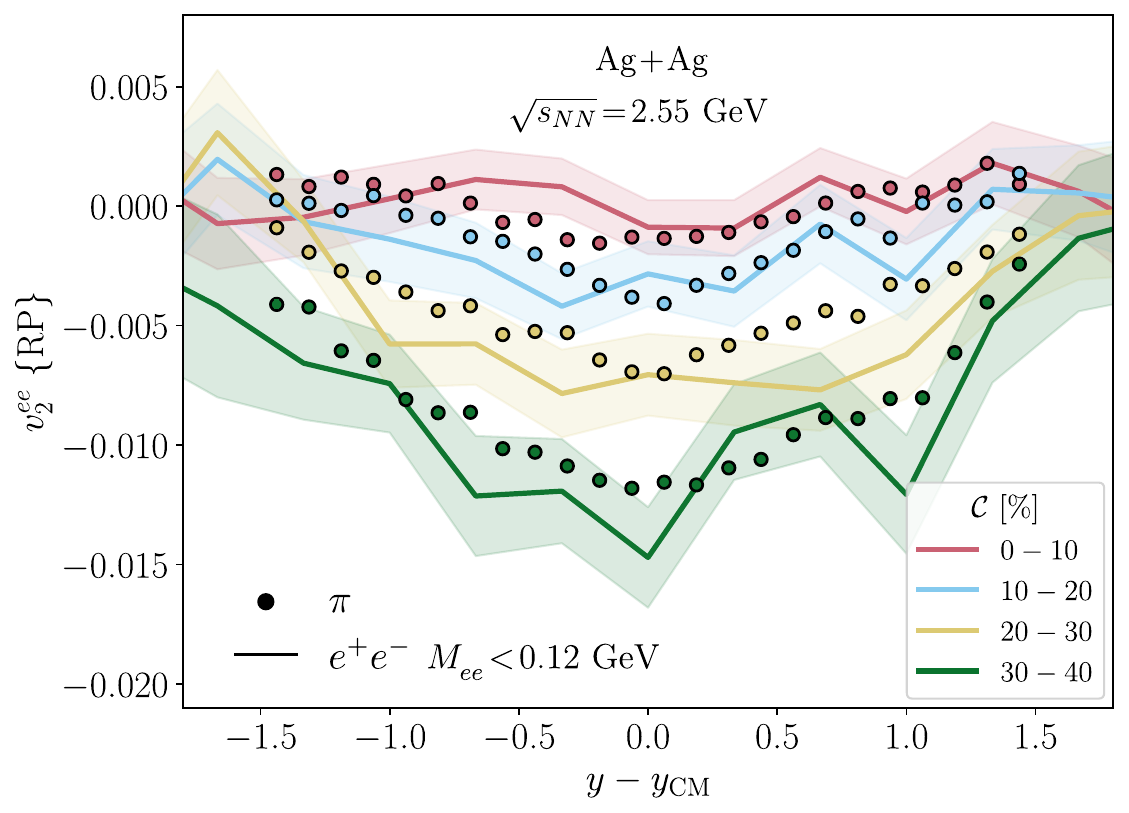}%
\includegraphics[width=0.333\linewidth]{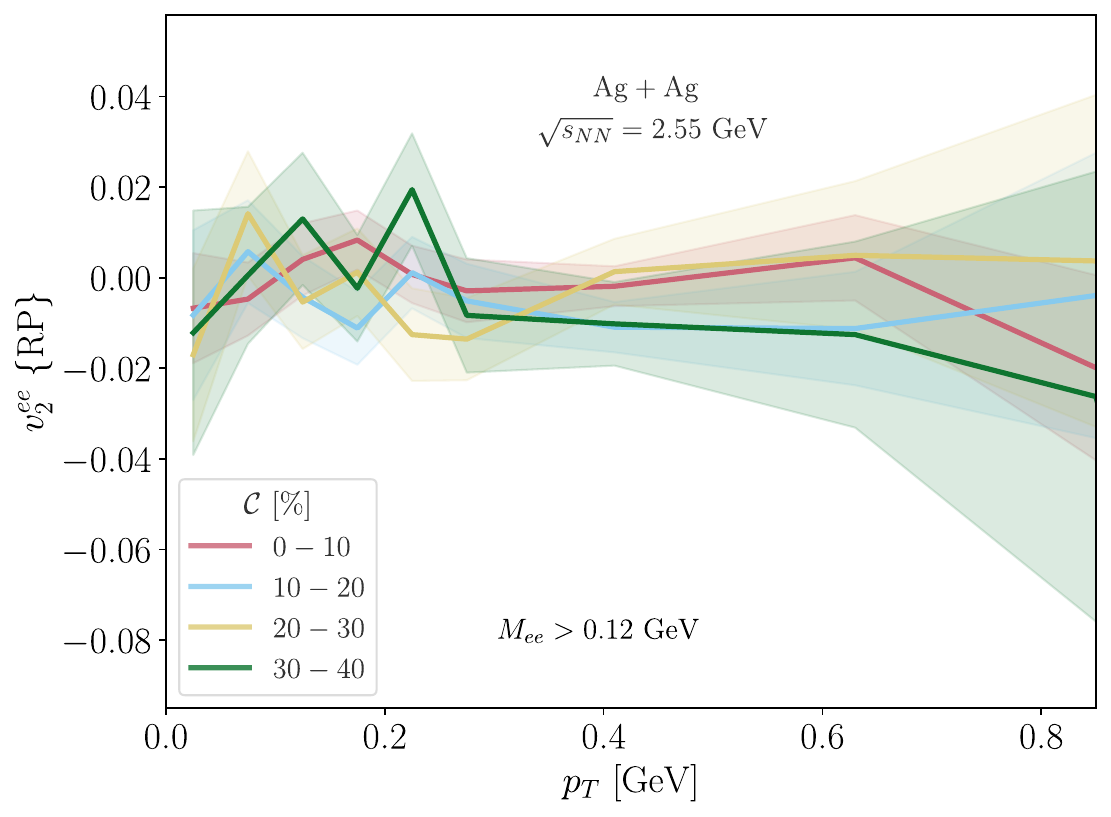}
\caption{Differential $v_2^{ee}$ as a function of the dilepton transverse momentum in the (left) pion and (right) resonance mass regions, and as a function of rapidity for low masses (middle).}\label{flow:fig:RP_masscut}
\end{figure}

We show predictions for the elliptic flow of dielectrons calculated with the reaction plane method as a function of $p_T$ and $y$ in \cref{flow:fig:RP_masscut}. We remark again that the events were evolved without nuclear potentials, which tend to increase the flow.

In the left and middle plots, we see that low-mass dielectrons have a flow consistent with that of pions, which serves as a sanity check, while on the middle plot the flow signal of the resonance region is again consistent with 0 for all centralities, as seen in \cref{flow:fig:RP_invmass}. 

\section{Transverse momentum dependence in SP}\label{sec:app:SP_pT}

Here we look at the dependence of the scalar product $v_2^{ee}$ on tranverse momentum for the two mass regimes:

\begin{figure}[h!]
\centering
\includegraphics[width=0.5\linewidth]{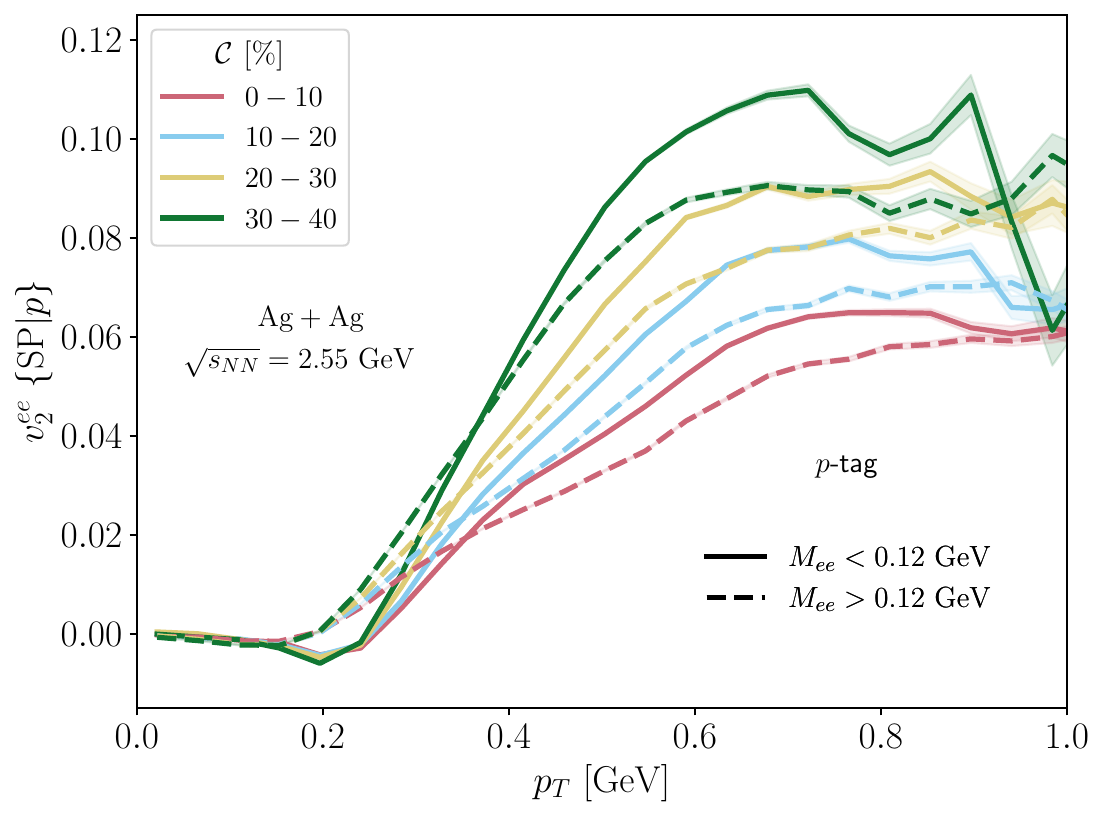}%
\includegraphics[width=0.5\linewidth]{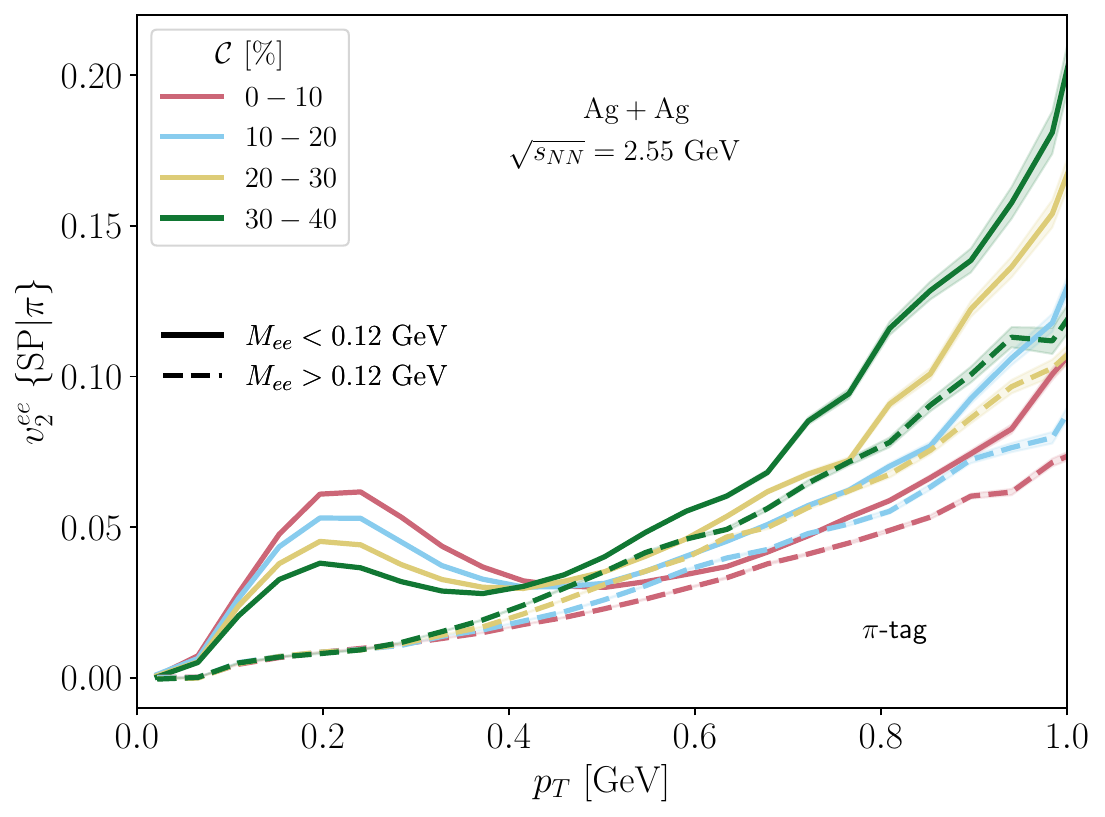}
\caption{Sources of proton-tagged dielectron $v_2$ with the scalar product method as a function of the transverse momentum for low invariant masses.}\label{flow:fig:SP_pT}
\end{figure}

We show the $p$-tagged flow in the left of \cref{flow:fig:SP_pT}. Generally, it is zero for low $p_T$ and increases linearly between $0.2$ and $0.5$ $\GeV$, when it saturates. The centrality dependence follows the expected trend, with stronger signals for more peripheral collisions.  In the right plot, the $\pi$-tag has a bump as low masses and low momentum, originating from the slow pions that are produced by decays of resonances already in the dilute stage. For both tags, the lower masses exhibit a larger signal at high $p_T$.

\bibliographystyle{apsrev4-1}
\bibliography{references}

\begin{thebibliography}{37}%
\makeatletter
\providecommand \@ifxundefined [1]{%
 \@ifx{#1\undefined}
}%
\providecommand \@ifnum [1]{%
 \ifnum #1\expandafter \@firstoftwo
 \else \expandafter \@secondoftwo
 \fi
}%
\providecommand \@ifx [1]{%
 \ifx #1\expandafter \@firstoftwo
 \else \expandafter \@secondoftwo
 \fi
}%
\providecommand \natexlab [1]{#1}%
\providecommand \enquote  [1]{``#1''}%
\providecommand \bibnamefont  [1]{#1}%
\providecommand \bibfnamefont [1]{#1}%
\providecommand \citenamefont [1]{#1}%
\providecommand \href@noop [0]{\@secondoftwo}%
\providecommand \href [0]{\begingroup \@sanitize@url \@href}%
\providecommand \@href[1]{\@@startlink{#1}\@@href}%
\providecommand \@@href[1]{\endgroup#1\@@endlink}%
\providecommand \@sanitize@url [0]{\catcode `\\12\catcode `\$12\catcode
  `\&12\catcode `\#12\catcode `\^12\catcode `\_12\catcode `\%12\relax}%
\providecommand \@@startlink[1]{}%
\providecommand \@@endlink[0]{}%
\providecommand \url  [0]{\begingroup\@sanitize@url \@url }%
\providecommand \@url [1]{\endgroup\@href {#1}{\urlprefix }}%
\providecommand \urlprefix  [0]{URL }%
\providecommand \Eprint [0]{\href }%
\providecommand \doibase [0]{http://dx.doi.org/}%
\providecommand \selectlanguage [0]{\@gobble}%
\providecommand \bibinfo  [0]{\@secondoftwo}%
\providecommand \bibfield  [0]{\@secondoftwo}%
\providecommand \translation [1]{[#1]}%
\providecommand \BibitemOpen [0]{}%
\providecommand \bibitemStop [0]{}%
\providecommand \bibitemNoStop [0]{.\EOS\space}%
\providecommand \EOS [0]{\spacefactor3000\relax}%
\providecommand \BibitemShut  [1]{\csname bibitem#1\endcsname}%
\let\auto@bib@innerbib\@empty
\bibitem [{\citenamefont {Ollitrault}(1992)}]{Ollitrault:1992bk}%
  \BibitemOpen
  \bibfield  {author} {\bibinfo {author} {\bibfnamefont {J.-Y.}\ \bibnamefont
  {Ollitrault}},\ }\href {\doibase 10.1103/PhysRevD.46.229} {\bibfield
  {journal} {\bibinfo  {journal} {Phys. Rev. D}\ }\textbf {\bibinfo {volume}
  {46}},\ \bibinfo {pages} {229} (\bibinfo {year} {1992})}\BibitemShut
  {NoStop}%
\bibitem [{\citenamefont {Luzum}\ and\ \citenamefont
  {Petersen}(2014)}]{Luzum:2013yya}%
  \BibitemOpen
  \bibfield  {author} {\bibinfo {author} {\bibfnamefont {M.}~\bibnamefont
  {Luzum}}\ and\ \bibinfo {author} {\bibfnamefont {H.}~\bibnamefont
  {Petersen}},\ }\href {\doibase 10.1088/0954-3899/41/6/063102} {\bibfield
  {journal} {\bibinfo  {journal} {J. Phys. G}\ }\textbf {\bibinfo {volume}
  {41}},\ \bibinfo {pages} {063102} (\bibinfo {year} {2014})},\ \Eprint
  {http://arxiv.org/abs/1312.5503} {arXiv:1312.5503 [nucl-th]} \BibitemShut
  {NoStop}%
\bibitem [{\citenamefont {Vujanovic}\ \emph {et~al.}(2020)\citenamefont
  {Vujanovic}, \citenamefont {Paquet}, \citenamefont {Shen}, \citenamefont
  {Denicol}, \citenamefont {Jeon}, \citenamefont {Gale},\ and\ \citenamefont
  {Heinz}}]{Vujanovic:2019yih}%
  \BibitemOpen
  \bibfield  {author} {\bibinfo {author} {\bibfnamefont {G.}~\bibnamefont
  {Vujanovic}}, \bibinfo {author} {\bibfnamefont {J.-F.}\ \bibnamefont
  {Paquet}}, \bibinfo {author} {\bibfnamefont {C.}~\bibnamefont {Shen}},
  \bibinfo {author} {\bibfnamefont {G.~S.}\ \bibnamefont {Denicol}}, \bibinfo
  {author} {\bibfnamefont {S.}~\bibnamefont {Jeon}}, \bibinfo {author}
  {\bibfnamefont {C.}~\bibnamefont {Gale}}, \ and\ \bibinfo {author}
  {\bibfnamefont {U.}~\bibnamefont {Heinz}},\ }\href {\doibase
  10.1103/PhysRevC.101.044904} {\bibfield  {journal} {\bibinfo  {journal}
  {Phys. Rev. C}\ }\textbf {\bibinfo {volume} {101}},\ \bibinfo {pages}
  {044904} (\bibinfo {year} {2020})},\ \Eprint
  {http://arxiv.org/abs/1903.05078} {arXiv:1903.05078 [nucl-th]} \BibitemShut
  {NoStop}%
\bibitem [{\citenamefont {Vujanovic}\ \emph {et~al.}(2016)\citenamefont
  {Vujanovic}, \citenamefont {Paquet}, \citenamefont {Denicol}, \citenamefont
  {Luzum}, \citenamefont {Jeon},\ and\ \citenamefont
  {Gale}}]{Vujanovic:2016anq}%
  \BibitemOpen
  \bibfield  {author} {\bibinfo {author} {\bibfnamefont {G.}~\bibnamefont
  {Vujanovic}}, \bibinfo {author} {\bibfnamefont {J.-F.}\ \bibnamefont
  {Paquet}}, \bibinfo {author} {\bibfnamefont {G.~S.}\ \bibnamefont {Denicol}},
  \bibinfo {author} {\bibfnamefont {M.}~\bibnamefont {Luzum}}, \bibinfo
  {author} {\bibfnamefont {S.}~\bibnamefont {Jeon}}, \ and\ \bibinfo {author}
  {\bibfnamefont {C.}~\bibnamefont {Gale}},\ }\href {\doibase
  10.1103/PhysRevC.94.014904} {\bibfield  {journal} {\bibinfo  {journal} {Phys.
  Rev. C}\ }\textbf {\bibinfo {volume} {94}},\ \bibinfo {pages} {014904}
  (\bibinfo {year} {2016})},\ \Eprint {http://arxiv.org/abs/1602.01455}
  {arXiv:1602.01455 [nucl-th]} \BibitemShut {NoStop}%
\bibitem [{\citenamefont {Vujanovic}\ \emph {et~al.}(2018)\citenamefont
  {Vujanovic}, \citenamefont {Denicol}, \citenamefont {Luzum}, \citenamefont
  {Jeon},\ and\ \citenamefont {Gale}}]{Vujanovic:2017psb}%
  \BibitemOpen
  \bibfield  {author} {\bibinfo {author} {\bibfnamefont {G.}~\bibnamefont
  {Vujanovic}}, \bibinfo {author} {\bibfnamefont {G.~S.}\ \bibnamefont
  {Denicol}}, \bibinfo {author} {\bibfnamefont {M.}~\bibnamefont {Luzum}},
  \bibinfo {author} {\bibfnamefont {S.}~\bibnamefont {Jeon}}, \ and\ \bibinfo
  {author} {\bibfnamefont {C.}~\bibnamefont {Gale}},\ }\href {\doibase
  10.1103/PhysRevC.98.014902} {\bibfield  {journal} {\bibinfo  {journal} {Phys.
  Rev. C}\ }\textbf {\bibinfo {volume} {98}},\ \bibinfo {pages} {014902}
  (\bibinfo {year} {2018})},\ \Eprint {http://arxiv.org/abs/1702.02941}
  {arXiv:1702.02941 [nucl-th]} \BibitemShut {NoStop}%
\bibitem [{\citenamefont {Reichert}\ \emph {et~al.}(2023)\citenamefont
  {Reichert}, \citenamefont {Savchuk}, \citenamefont {Kittiratpattana},
  \citenamefont {Li}, \citenamefont {Steinheimer}, \citenamefont {Gorenstein},\
  and\ \citenamefont {Bleicher}}]{Reichert:2023eev}%
  \BibitemOpen
  \bibfield  {author} {\bibinfo {author} {\bibfnamefont {T.}~\bibnamefont
  {Reichert}}, \bibinfo {author} {\bibfnamefont {O.}~\bibnamefont {Savchuk}},
  \bibinfo {author} {\bibfnamefont {A.}~\bibnamefont {Kittiratpattana}},
  \bibinfo {author} {\bibfnamefont {P.}~\bibnamefont {Li}}, \bibinfo {author}
  {\bibfnamefont {J.}~\bibnamefont {Steinheimer}}, \bibinfo {author}
  {\bibfnamefont {M.}~\bibnamefont {Gorenstein}}, \ and\ \bibinfo {author}
  {\bibfnamefont {M.}~\bibnamefont {Bleicher}},\ }\href {\doibase
  10.1016/j.physletb.2023.137947} {\bibfield  {journal} {\bibinfo  {journal}
  {Phys. Lett. B}\ }\textbf {\bibinfo {volume} {841}},\ \bibinfo {pages}
  {137947} (\bibinfo {year} {2023})},\ \Eprint
  {http://arxiv.org/abs/2302.13919} {arXiv:2302.13919 [nucl-th]} \BibitemShut
  {NoStop}%
\bibitem [{\citenamefont {Stock}(2004)}]{Stock:2004iim}%
  \BibitemOpen
  \bibfield  {author} {\bibinfo {author} {\bibfnamefont {R.}~\bibnamefont
  {Stock}},\ }\href {\doibase 10.1088/0954-3899/30/8/001} {\bibfield  {journal}
  {\bibinfo  {journal} {J. Phys. G}\ }\textbf {\bibinfo {volume} {30}},\
  \bibinfo {pages} {S633} (\bibinfo {year} {2004})},\ \Eprint
  {http://arxiv.org/abs/nucl-ex/0405007} {arXiv:nucl-ex/0405007} \BibitemShut
  {NoStop}%
\bibitem [{\citenamefont {Agakishiev}\ \emph {et~al.}(2009)\citenamefont
  {Agakishiev} \emph {et~al.}}]{HADES:2009aat}%
  \BibitemOpen
  \bibfield  {author} {\bibinfo {author} {\bibfnamefont {G.}~\bibnamefont
  {Agakishiev}} \emph {et~al.} (\bibinfo {collaboration} {HADES}),\ }\href
  {\doibase 10.1140/epja/i2009-10807-5} {\bibfield  {journal} {\bibinfo
  {journal} {Eur. Phys. J. A}\ }\textbf {\bibinfo {volume} {41}},\ \bibinfo
  {pages} {243} (\bibinfo {year} {2009})},\ \Eprint
  {http://arxiv.org/abs/0902.3478} {arXiv:0902.3478 [nucl-ex]} \BibitemShut
  {NoStop}%
\bibitem [{\citenamefont {Abdallah}\ \emph {et~al.}(2022)\citenamefont
  {Abdallah} \emph {et~al.}}]{STAR:2022gki}%
  \BibitemOpen
  \bibfield  {author} {\bibinfo {author} {\bibfnamefont {M.}~\bibnamefont
  {Abdallah}} \emph {et~al.} (\bibinfo {collaboration} {STAR}),\ }\href
  {\doibase 10.1103/PhysRevLett.129.252301} {\bibfield  {journal} {\bibinfo
  {journal} {Phys. Rev. Lett.}\ }\textbf {\bibinfo {volume} {129}},\ \bibinfo
  {pages} {252301} (\bibinfo {year} {2022})},\ \Eprint
  {http://arxiv.org/abs/2201.10365} {arXiv:2201.10365 [nucl-ex]} \BibitemShut
  {NoStop}%
\bibitem [{The sPHENIX Collab.()}]{sPHENIX:2024flow}%
  \BibitemOpen
  The sPHENIX Collab.,\ \href@noop {} {\enquote {\bibinfo {title} {{Measurement
  of $\pi^0$ $v_2$ in Au+Au Collisions at 200 GeV with the sPHENIX
  Detector}},}\ } (\bibinfo {year} {2024})\BibitemShut {NoStop}%
\bibitem [{\citenamefont {Khachatryan}\ \emph {et~al.}(2015)\citenamefont
  {Khachatryan} \emph {et~al.}}]{CMS:2015xmx}%
  \BibitemOpen
  \bibfield  {author} {\bibinfo {author} {\bibfnamefont {V.}~\bibnamefont
  {Khachatryan}} \emph {et~al.} (\bibinfo {collaboration} {CMS}),\ }\href
  {\doibase 10.1103/PhysRevC.92.034911} {\bibfield  {journal} {\bibinfo
  {journal} {Phys. Rev. C}\ }\textbf {\bibinfo {volume} {92}},\ \bibinfo
  {pages} {034911} (\bibinfo {year} {2015})},\ \Eprint
  {http://arxiv.org/abs/1503.01692} {arXiv:1503.01692 [nucl-ex]} \BibitemShut
  {NoStop}%
\bibitem [{\citenamefont {Acharya}\ \emph {et~al.}(2023)\citenamefont {Acharya}
  \emph {et~al.}}]{ALICE:2022zks}%
  \BibitemOpen
  \bibfield  {author} {\bibinfo {author} {\bibfnamefont {S.}~\bibnamefont
  {Acharya}} \emph {et~al.} (\bibinfo {collaboration} {ALICE}),\ }\href
  {\doibase 10.1007/JHEP05(2023)243} {\bibfield  {journal} {\bibinfo  {journal}
  {JHEP}\ }\textbf {\bibinfo {volume} {05}},\ \bibinfo {pages} {243} (\bibinfo
  {year} {2023})},\ \Eprint {http://arxiv.org/abs/2206.04587} {arXiv:2206.04587
  [nucl-ex]} \BibitemShut {NoStop}%
\bibitem [{\citenamefont {Voloshin}\ \emph {et~al.}(2010)\citenamefont
  {Voloshin}, \citenamefont {Poskanzer},\ and\ \citenamefont
  {Snellings}}]{Voloshin:2008dg}%
  \BibitemOpen
  \bibfield  {author} {\bibinfo {author} {\bibfnamefont {S.~A.}\ \bibnamefont
  {Voloshin}}, \bibinfo {author} {\bibfnamefont {A.~M.}\ \bibnamefont
  {Poskanzer}}, \ and\ \bibinfo {author} {\bibfnamefont {R.}~\bibnamefont
  {Snellings}},\ }\href {\doibase 10.1007/978-3-642-01539-7_10} {\bibfield
  {journal} {\bibinfo  {journal} {Landolt-Bornstein}\ }\textbf {\bibinfo
  {volume} {23}},\ \bibinfo {pages} {293} (\bibinfo {year} {2010})},\ \Eprint
  {http://arxiv.org/abs/0809.2949} {arXiv:0809.2949 [nucl-ex]} \BibitemShut
  {NoStop}%
\bibitem [{\citenamefont {Borghini}\ \emph {et~al.}(2001)\citenamefont
  {Borghini}, \citenamefont {Dinh},\ and\ \citenamefont
  {Ollitrault}}]{Borghini:2001vi}%
  \BibitemOpen
  \bibfield  {author} {\bibinfo {author} {\bibfnamefont {N.}~\bibnamefont
  {Borghini}}, \bibinfo {author} {\bibfnamefont {P.~M.}\ \bibnamefont {Dinh}},
  \ and\ \bibinfo {author} {\bibfnamefont {J.-Y.}\ \bibnamefont {Ollitrault}},\
  }\href {\doibase 10.1103/PhysRevC.64.054901} {\bibfield  {journal} {\bibinfo
  {journal} {Phys. Rev. C}\ }\textbf {\bibinfo {volume} {64}},\ \bibinfo
  {pages} {054901} (\bibinfo {year} {2001})},\ \Eprint
  {http://arxiv.org/abs/nucl-th/0105040} {arXiv:nucl-th/0105040} \BibitemShut
  {NoStop}%
\bibitem [{\citenamefont {Kardan}(2019)}]{Kardan:2018hna}%
  \BibitemOpen
  \bibfield  {author} {\bibinfo {author} {\bibfnamefont {B.}~\bibnamefont
  {Kardan}} (\bibinfo {collaboration} {HADES}),\ }\href {\doibase
  10.1016/j.nuclphysa.2018.09.061} {\bibfield  {journal} {\bibinfo  {journal}
  {Nucl. Phys. A}\ }\textbf {\bibinfo {volume} {982}},\ \bibinfo {pages} {431}
  (\bibinfo {year} {2019})},\ \Eprint {http://arxiv.org/abs/1809.07821}
  {arXiv:1809.07821 [nucl-ex]} \BibitemShut {NoStop}%
\bibitem [{\citenamefont {Adamczewski-Musch}\ \emph {et~al.}(2020)\citenamefont
  {Adamczewski-Musch} \emph {et~al.}}]{HADES:2020lob}%
  \BibitemOpen
  \bibfield  {author} {\bibinfo {author} {\bibfnamefont {J.}~\bibnamefont
  {Adamczewski-Musch}} \emph {et~al.} (\bibinfo {collaboration} {HADES}),\
  }\href {\doibase 10.1103/PhysRevLett.125.262301} {\bibfield  {journal}
  {\bibinfo  {journal} {Phys. Rev. Lett.}\ }\textbf {\bibinfo {volume} {125}},\
  \bibinfo {pages} {262301} (\bibinfo {year} {2020})},\ \Eprint
  {http://arxiv.org/abs/2005.12217} {arXiv:2005.12217 [nucl-ex]} \BibitemShut
  {NoStop}%
\bibitem [{\citenamefont {Galatyuk}(2020)}]{Galatyuk:2020lvg}%
  \BibitemOpen
  \bibfield  {author} {\bibinfo {author} {\bibfnamefont {T.}~\bibnamefont
  {Galatyuk}} (\bibinfo {collaboration} {HADES}),\ }\href {\doibase
  10.7566/JPSCP.32.010079} {\bibfield  {journal} {\bibinfo  {journal} {JPS
  Conf. Proc.}\ }\textbf {\bibinfo {volume} {32}},\ \bibinfo {pages} {010079}
  (\bibinfo {year} {2020})}\BibitemShut {NoStop}%
\bibitem [{\citenamefont {Schild}(2024)}]{Schild:2024ywo}%
  \BibitemOpen
  \bibfield  {author} {\bibinfo {author} {\bibfnamefont {N.}~\bibnamefont
  {Schild}} (\bibinfo {collaboration} {HADES}),\ }\href {\doibase
  10.22323/1.438.0072} {\bibfield  {journal} {\bibinfo  {journal} {PoS}\
  }\textbf {\bibinfo {volume} {HardProbes2023}},\ \bibinfo {pages} {072}
  (\bibinfo {year} {2024})}\BibitemShut {NoStop}%
\bibitem [{\citenamefont {Galatyuk}\ and\ \citenamefont
  {Schild}()}]{TalkWithGalatyukSchild}%
  \BibitemOpen
  \bibfield  {author} {\bibinfo {author} {\bibfnamefont {T.}~\bibnamefont
  {Galatyuk}}\ and\ \bibinfo {author} {\bibfnamefont {N.}~\bibnamefont
  {Schild}},\ }\href@noop {} {}\bibinfo {howpublished} {private
  communication}\BibitemShut {NoStop}%
\bibitem [{\citenamefont {Weil}\ \emph {et~al.}(2016)\citenamefont {Weil} \emph
  {et~al.}}]{SMASH:2016zqf}%
  \BibitemOpen
  \bibfield  {author} {\bibinfo {author} {\bibfnamefont {J.}~\bibnamefont
  {Weil}} \emph {et~al.} (\bibinfo {collaboration} {SMASH}),\ }\href {\doibase
  10.1103/PhysRevC.94.054905} {\bibfield  {journal} {\bibinfo  {journal} {Phys.
  Rev. C}\ }\textbf {\bibinfo {volume} {94}},\ \bibinfo {pages} {054905}
  (\bibinfo {year} {2016})},\ \Eprint {http://arxiv.org/abs/1606.06642}
  {arXiv:1606.06642 [nucl-th]} \BibitemShut {NoStop}%
\bibitem [{\citenamefont {Staudenmaier}\ \emph {et~al.}(2018)\citenamefont
  {Staudenmaier}, \citenamefont {Weil}, \citenamefont {Steinberg},
  \citenamefont {Endres},\ and\ \citenamefont
  {Petersen}}]{Staudenmaier:2017vtq}%
  \BibitemOpen
  \bibfield  {author} {\bibinfo {author} {\bibfnamefont {J.}~\bibnamefont
  {Staudenmaier}}, \bibinfo {author} {\bibfnamefont {J.}~\bibnamefont {Weil}},
  \bibinfo {author} {\bibfnamefont {V.}~\bibnamefont {Steinberg}}, \bibinfo
  {author} {\bibfnamefont {S.}~\bibnamefont {Endres}}, \ and\ \bibinfo {author}
  {\bibfnamefont {H.}~\bibnamefont {Petersen}} (\bibinfo {collaboration}
  {SMASH}),\ }\href {\doibase 10.1103/PhysRevC.98.054908} {\bibfield  {journal}
  {\bibinfo  {journal} {Phys. Rev. C}\ }\textbf {\bibinfo {volume} {98}},\
  \bibinfo {pages} {054908} (\bibinfo {year} {2018})},\ \Eprint
  {http://arxiv.org/abs/1711.10297} {arXiv:1711.10297 [nucl-th]} \BibitemShut
  {NoStop}%
\bibitem [{\citenamefont {Tanabashi}\ \emph {et~al.}(2018)\citenamefont
  {Tanabashi} \emph {et~al.}}]{ParticleDataGroup:2018ovx}%
  \BibitemOpen
  \bibfield  {author} {\bibinfo {author} {\bibfnamefont {M.}~\bibnamefont
  {Tanabashi}} \emph {et~al.} (\bibinfo {collaboration} {Particle Data
  Group}),\ }\href {\doibase 10.1103/PhysRevD.98.030001} {\bibfield  {journal}
  {\bibinfo  {journal} {Phys. Rev. D}\ }\textbf {\bibinfo {volume} {98}},\
  \bibinfo {pages} {030001} (\bibinfo {year} {2018})}\BibitemShut {NoStop}%
\bibitem [{\citenamefont {Petersen}\ \emph {et~al.}(2019)\citenamefont
  {Petersen}, \citenamefont {Oliinychenko}, \citenamefont {Mayer},
  \citenamefont {Staudenmaier},\ and\ \citenamefont {Ryu}}]{Petersen:2018jag}%
  \BibitemOpen
  \bibfield  {author} {\bibinfo {author} {\bibfnamefont {H.}~\bibnamefont
  {Petersen}}, \bibinfo {author} {\bibfnamefont {D.}~\bibnamefont
  {Oliinychenko}}, \bibinfo {author} {\bibfnamefont {M.}~\bibnamefont {Mayer}},
  \bibinfo {author} {\bibfnamefont {J.}~\bibnamefont {Staudenmaier}}, \ and\
  \bibinfo {author} {\bibfnamefont {S.}~\bibnamefont {Ryu}},\ }\href {\doibase
  10.1016/j.nuclphysa.2018.08.008} {\bibfield  {journal} {\bibinfo  {journal}
  {Nucl. Phys. A}\ }\textbf {\bibinfo {volume} {982}},\ \bibinfo {pages} {399}
  (\bibinfo {year} {2019})},\ \Eprint {http://arxiv.org/abs/1808.06832}
  {arXiv:1808.06832 [nucl-th]} \BibitemShut {NoStop}%
\bibitem [{\citenamefont {Mohs}\ \emph {et~al.}(2022)\citenamefont {Mohs},
  \citenamefont {Ege}, \citenamefont {Elfner},\ and\ \citenamefont
  {Mayer}}]{Mohs:2020awg}%
  \BibitemOpen
  \bibfield  {author} {\bibinfo {author} {\bibfnamefont {J.}~\bibnamefont
  {Mohs}}, \bibinfo {author} {\bibfnamefont {M.}~\bibnamefont {Ege}}, \bibinfo
  {author} {\bibfnamefont {H.}~\bibnamefont {Elfner}}, \ and\ \bibinfo {author}
  {\bibfnamefont {M.}~\bibnamefont {Mayer}} (\bibinfo {collaboration}
  {SMASH}),\ }\href {\doibase 10.1103/PhysRevC.105.034906} {\bibfield
  {journal} {\bibinfo  {journal} {Phys. Rev. C}\ }\textbf {\bibinfo {volume}
  {105}},\ \bibinfo {pages} {034906} (\bibinfo {year} {2022})},\ \Eprint
  {http://arxiv.org/abs/2012.11454} {arXiv:2012.11454 [nucl-th]} \BibitemShut
  {NoStop}%
\bibitem [{\citenamefont {Tarasovi\v{c}ov\'a}\ \emph
  {et~al.}(2024)\citenamefont {Tarasovi\v{c}ov\'a}, \citenamefont {Mohs},
  \citenamefont {Andronic}, \citenamefont {Elfner},\ and\ \citenamefont
  {Kampert}}]{Tarasovicova:2024isp}%
  \BibitemOpen
  \bibfield  {author} {\bibinfo {author} {\bibfnamefont {L.~A.}\ \bibnamefont
  {Tarasovi\v{c}ov\'a}}, \bibinfo {author} {\bibfnamefont {J.}~\bibnamefont
  {Mohs}}, \bibinfo {author} {\bibfnamefont {A.}~\bibnamefont {Andronic}},
  \bibinfo {author} {\bibfnamefont {H.}~\bibnamefont {Elfner}}, \ and\ \bibinfo
  {author} {\bibfnamefont {K.-H.}\ \bibnamefont {Kampert}},\ }\href@noop {} {\
  (\bibinfo {year} {2024})},\ \Eprint {http://arxiv.org/abs/2405.09889}
  {arXiv:2405.09889 [nucl-th]} \BibitemShut {NoStop}%
\bibitem [{\citenamefont {Heinz}\ and\ \citenamefont
  {Lee}(1992)}]{Heinz:1991fn}%
  \BibitemOpen
  \bibfield  {author} {\bibinfo {author} {\bibfnamefont {U.~W.}\ \bibnamefont
  {Heinz}}\ and\ \bibinfo {author} {\bibfnamefont {K.~S.}\ \bibnamefont
  {Lee}},\ }\href {\doibase 10.1016/0375-9474(92)90606-K} {\bibfield  {journal}
  {\bibinfo  {journal} {Nucl. Phys. A}\ }\textbf {\bibinfo {volume} {544}},\
  \bibinfo {pages} {503} (\bibinfo {year} {1992})}\BibitemShut {NoStop}%
\bibitem [{\citenamefont {Hirayama}\ \emph {et~al.}(2023)\citenamefont
  {Hirayama}, \citenamefont {Staudenmaier},\ and\ \citenamefont
  {Elfner}}]{Hirayama:2022rur}%
  \BibitemOpen
  \bibfield  {author} {\bibinfo {author} {\bibfnamefont {R.}~\bibnamefont
  {Hirayama}}, \bibinfo {author} {\bibfnamefont {J.}~\bibnamefont
  {Staudenmaier}}, \ and\ \bibinfo {author} {\bibfnamefont {H.}~\bibnamefont
  {Elfner}} (\bibinfo {collaboration} {SMASH}),\ }\href {\doibase
  10.1103/PhysRevC.107.025208} {\bibfield  {journal} {\bibinfo  {journal}
  {Phys. Rev. C}\ }\textbf {\bibinfo {volume} {107}},\ \bibinfo {pages}
  {025208} (\bibinfo {year} {2023})},\ \Eprint
  {http://arxiv.org/abs/2206.15166} {arXiv:2206.15166 [hep-ph]} \BibitemShut
  {NoStop}%
\bibitem [{\citenamefont {Rapp}\ and\ \citenamefont
  {Wambach}(1999)}]{Rapp:1999us}%
  \BibitemOpen
  \bibfield  {author} {\bibinfo {author} {\bibfnamefont {R.}~\bibnamefont
  {Rapp}}\ and\ \bibinfo {author} {\bibfnamefont {J.}~\bibnamefont {Wambach}},\
  }\href {\doibase 10.1007/s100500050364} {\bibfield  {journal} {\bibinfo
  {journal} {Eur. Phys. J. A}\ }\textbf {\bibinfo {volume} {6}},\ \bibinfo
  {pages} {415} (\bibinfo {year} {1999})},\ \Eprint
  {http://arxiv.org/abs/hep-ph/9907502} {arXiv:hep-ph/9907502} \BibitemShut
  {NoStop}%
\bibitem [{\citenamefont {Rapp}\ \emph {et~al.}(2010)\citenamefont {Rapp},
  \citenamefont {Wambach},\ and\ \citenamefont {van Hees}}]{Rapp:2009yu}%
  \BibitemOpen
  \bibfield  {author} {\bibinfo {author} {\bibfnamefont {R.}~\bibnamefont
  {Rapp}}, \bibinfo {author} {\bibfnamefont {J.}~\bibnamefont {Wambach}}, \
  and\ \bibinfo {author} {\bibfnamefont {H.}~\bibnamefont {van Hees}},\ }\href
  {\doibase 10.1007/978-3-642-01539-7_6} {\bibfield  {journal} {\bibinfo
  {journal} {Landolt-Bornstein}\ }\textbf {\bibinfo {volume} {23}},\ \bibinfo
  {pages} {134} (\bibinfo {year} {2010})},\ \Eprint
  {http://arxiv.org/abs/0901.3289} {arXiv:0901.3289 [hep-ph]} \BibitemShut
  {NoStop}%
\bibitem [{\citenamefont {Endres}\ \emph
  {et~al.}(2015{\natexlab{a}})\citenamefont {Endres}, \citenamefont {van Hees},
  \citenamefont {Weil},\ and\ \citenamefont {Bleicher}}]{Endres:2014zua}%
  \BibitemOpen
  \bibfield  {author} {\bibinfo {author} {\bibfnamefont {S.}~\bibnamefont
  {Endres}}, \bibinfo {author} {\bibfnamefont {H.}~\bibnamefont {van Hees}},
  \bibinfo {author} {\bibfnamefont {J.}~\bibnamefont {Weil}}, \ and\ \bibinfo
  {author} {\bibfnamefont {M.}~\bibnamefont {Bleicher}},\ }\href {\doibase
  10.1103/PhysRevC.91.054911} {\bibfield  {journal} {\bibinfo  {journal} {Phys.
  Rev. C}\ }\textbf {\bibinfo {volume} {91}},\ \bibinfo {pages} {054911}
  (\bibinfo {year} {2015}{\natexlab{a}})},\ \Eprint
  {http://arxiv.org/abs/1412.1965} {arXiv:1412.1965 [nucl-th]} \BibitemShut
  {NoStop}%
\bibitem [{\citenamefont {Endres}\ \emph
  {et~al.}(2015{\natexlab{b}})\citenamefont {Endres}, \citenamefont {van Hees},
  \citenamefont {Weil},\ and\ \citenamefont {Bleicher}}]{Endres:2015fna}%
  \BibitemOpen
  \bibfield  {author} {\bibinfo {author} {\bibfnamefont {S.}~\bibnamefont
  {Endres}}, \bibinfo {author} {\bibfnamefont {H.}~\bibnamefont {van Hees}},
  \bibinfo {author} {\bibfnamefont {J.}~\bibnamefont {Weil}}, \ and\ \bibinfo
  {author} {\bibfnamefont {M.}~\bibnamefont {Bleicher}},\ }\href {\doibase
  10.1103/PhysRevC.92.014911} {\bibfield  {journal} {\bibinfo  {journal} {Phys.
  Rev. C}\ }\textbf {\bibinfo {volume} {92}},\ \bibinfo {pages} {014911}
  (\bibinfo {year} {2015}{\natexlab{b}})},\ \Eprint
  {http://arxiv.org/abs/1505.06131} {arXiv:1505.06131 [nucl-th]} \BibitemShut
  {NoStop}%
\bibitem [{\citenamefont {Galatyuk}\ \emph {et~al.}(2016)\citenamefont
  {Galatyuk}, \citenamefont {Hohler}, \citenamefont {Rapp}, \citenamefont
  {Seck},\ and\ \citenamefont {Stroth}}]{Galatyuk:2015pkq}%
  \BibitemOpen
  \bibfield  {author} {\bibinfo {author} {\bibfnamefont {T.}~\bibnamefont
  {Galatyuk}}, \bibinfo {author} {\bibfnamefont {P.~M.}\ \bibnamefont
  {Hohler}}, \bibinfo {author} {\bibfnamefont {R.}~\bibnamefont {Rapp}},
  \bibinfo {author} {\bibfnamefont {F.}~\bibnamefont {Seck}}, \ and\ \bibinfo
  {author} {\bibfnamefont {J.}~\bibnamefont {Stroth}},\ }\href {\doibase
  10.1140/epja/i2016-16131-1} {\bibfield  {journal} {\bibinfo  {journal} {Eur.
  Phys. J. A}\ }\textbf {\bibinfo {volume} {52}},\ \bibinfo {pages} {131}
  (\bibinfo {year} {2016})},\ \Eprint {http://arxiv.org/abs/1512.08688}
  {arXiv:1512.08688 [nucl-th]} \BibitemShut {NoStop}%
\bibitem [{\citenamefont {Seck}\ \emph {et~al.}(2022)\citenamefont {Seck},
  \citenamefont {Galatyuk}, \citenamefont {Mukherjee}, \citenamefont {Rapp},
  \citenamefont {Steinheimer}, \citenamefont {Stroth},\ and\ \citenamefont
  {Wiest}}]{Seck:2020qbx}%
  \BibitemOpen
  \bibfield  {author} {\bibinfo {author} {\bibfnamefont {F.}~\bibnamefont
  {Seck}}, \bibinfo {author} {\bibfnamefont {T.}~\bibnamefont {Galatyuk}},
  \bibinfo {author} {\bibfnamefont {A.}~\bibnamefont {Mukherjee}}, \bibinfo
  {author} {\bibfnamefont {R.}~\bibnamefont {Rapp}}, \bibinfo {author}
  {\bibfnamefont {J.}~\bibnamefont {Steinheimer}}, \bibinfo {author}
  {\bibfnamefont {J.}~\bibnamefont {Stroth}}, \ and\ \bibinfo {author}
  {\bibfnamefont {M.}~\bibnamefont {Wiest}},\ }\href {\doibase
  10.1103/PhysRevC.106.014904} {\bibfield  {journal} {\bibinfo  {journal}
  {Phys. Rev. C}\ }\textbf {\bibinfo {volume} {106}},\ \bibinfo {pages}
  {014904} (\bibinfo {year} {2022})},\ \Eprint
  {http://arxiv.org/abs/2010.04614} {arXiv:2010.04614 [nucl-th]} \BibitemShut
  {NoStop}%
\bibitem [{\citenamefont {Voloshin}\ and\ \citenamefont
  {Zhang}(1996)}]{Voloshin:1994mz}%
  \BibitemOpen
  \bibfield  {author} {\bibinfo {author} {\bibfnamefont {S.}~\bibnamefont
  {Voloshin}}\ and\ \bibinfo {author} {\bibfnamefont {Y.}~\bibnamefont
  {Zhang}},\ }\href {\doibase 10.1007/s002880050141} {\bibfield  {journal}
  {\bibinfo  {journal} {Z. Phys. C}\ }\textbf {\bibinfo {volume} {70}},\
  \bibinfo {pages} {665} (\bibinfo {year} {1996})},\ \Eprint
  {http://arxiv.org/abs/hep-ph/9407282} {arXiv:hep-ph/9407282} \BibitemShut
  {NoStop}%
\bibitem [{\citenamefont {Prozorov}(2024)}]{Prozorov:2024wto}%
  \BibitemOpen
  \bibfield  {author} {\bibinfo {author} {\bibfnamefont {A.}~\bibnamefont
  {Prozorov}} (\bibinfo {collaboration} {HADES}),\ }\href {\doibase
  10.1051/epjconf/202429104001} {\bibfield  {journal} {\bibinfo  {journal} {EPJ
  Web Conf.}\ }\textbf {\bibinfo {volume} {291}},\ \bibinfo {pages} {04001}
  (\bibinfo {year} {2024})}\BibitemShut {NoStop}%
\bibitem [{\citenamefont {Paquet}\ \emph {et~al.}(2016)\citenamefont {Paquet},
  \citenamefont {Shen}, \citenamefont {Denicol}, \citenamefont {Luzum},
  \citenamefont {Schenke}, \citenamefont {Jeon},\ and\ \citenamefont
  {Gale}}]{Paquet:2015lta}%
  \BibitemOpen
  \bibfield  {author} {\bibinfo {author} {\bibfnamefont {J.-F.}\ \bibnamefont
  {Paquet}}, \bibinfo {author} {\bibfnamefont {C.}~\bibnamefont {Shen}},
  \bibinfo {author} {\bibfnamefont {G.~S.}\ \bibnamefont {Denicol}}, \bibinfo
  {author} {\bibfnamefont {M.}~\bibnamefont {Luzum}}, \bibinfo {author}
  {\bibfnamefont {B.}~\bibnamefont {Schenke}}, \bibinfo {author} {\bibfnamefont
  {S.}~\bibnamefont {Jeon}}, \ and\ \bibinfo {author} {\bibfnamefont
  {C.}~\bibnamefont {Gale}},\ }\href {\doibase 10.1103/PhysRevC.93.044906}
  {\bibfield  {journal} {\bibinfo  {journal} {Phys. Rev. C}\ }\textbf {\bibinfo
  {volume} {93}},\ \bibinfo {pages} {044906} (\bibinfo {year} {2016})},\
  \Eprint {http://arxiv.org/abs/1509.06738} {arXiv:1509.06738 [hep-ph]}
  \BibitemShut {NoStop}%
\bibitem [{\citenamefont {Miskowiec}(2001)}]{nuclearoverlap}%
  \BibitemOpen
  \bibfield  {author} {\bibinfo {author} {\bibfnamefont {D.}~\bibnamefont
  {Miskowiec}},\ }\href {http://web-docs.gsi.de/~misko/overlap/interface.html}
  {\enquote {\bibinfo {title} {Web interface for a nuclear overlap calculation
  code},}\ } (\bibinfo {year} {2001})\BibitemShut {NoStop}%
\end{thebibliography}%

\end{document}